\documentclass[quantumrep,article,accept,pdftex,oneauthor]{Definitions/mdpi} 

\firstpage{1} 
\pubvolume{1}
\issuenum{1}
\articlenumber{0}
\pubyear{2024}
\copyrightyear{2024}
\externaleditor{Academic Editors: Viktor Dodonov, Margarita A. Man’ko, Salomon S Mizrahi, Luis L. Sánchez-Soto}
\datereceived{23 May 2024 } 
\daterevised{15 June 2024 } % Comment out if no revised date
\dateaccepted{16 June 2024 } 
\datepublished{ } 
\hreflink{https://doi.org/} % If needed use \linebreak
\usepackage{amsthm}
\usepackage{amsmath,amssymb,amsfonts,braket}
\usepackage{amsfonts,braket}
\usepackage{graphicx}       
\usepackage{comment}
\usepackage[labelformat=simple]{subcaption}

\DeclareCaptionLabelFormat{subcaptionlabel}{\normalfont(\textbf{#2}\normalfont)}
\usepackage{custom_shortcuts}

\Title{Eliminating the Second-Order Time Dependence from the Time Dependent Schr\"{o}dinger Equation Using Recursive Fourier~Transforms}%Attention: title altered

\TitleCitation{Eliminating the Second-Order Time Dependence from the Time Dependent Schr\"{o}dinger Equation Using Recursive Fourier Transforms}

\Author{Sky Nelson-Isaacs %MDPI: Please carefully check the accuracy of names and affiliations.
 \orcidA{}}

\AuthorNames{Sky Nelson-Isaacs}

\AuthorCitation{Nelson-Isaacs, S.}
\address[1]{Independent Scholar, El Cerrito,
 CA 94530, USA; theskyband@gmail.com}

\abstract{A strategy is developed for writing the time-dependent Schr\"{o}dinger Equation (TDSE), and more generally the Dyson Series, as a convolution equation using recursive Fourier transforms, thereby decoupling the second-order integral from the first without using the time ordering operator. The energy distribution is calculated for a number of standard perturbation theory examples at first- and second-order. Possible applications include characterization of photonic spectra for bosonic sampling and four-wave mixing in quantum computation and Bardeen tunneling amplitude in quantum mechanics.
}


\begin{document}

\section{Introduction}

The Time-Dependent Schrödinger Equation (TDSE), although~unsolvable in exact terms, is often approached through various perturbative methodologies, such as those pioneered by Rayleigh--Schrödinger~\cite{Schrod1926e}, Dirac~\cite{DIRAC1930}, Dyson~\cite{DYSON1952}, Lippmann--Schwinger~\cite{SCHWARTZ2014}, WKB~\cite{WALKER1994}, Feynman~\cite{FEYNMAN1948,FEYNMAN1948b,FEYNMAN1948c}, and~others~\cite{GRIFFITHS2018}. Under~a weak time-dependent perturbation, the TDSE solution can be written in terms of the known eigenvectors of the time independent Schrödinger Equation (TISE), which results in a Dyson series---an infinite recursion of coupled time integrals. In~field theory, it is standard to use the time-ordering operator for decoupling the~integrals.

Using the observation that these coupled integrals can be written as repeated Fourier transforms and their inverse, along with appropriate phase factors, a~method is developed for decoupling the integrals at second~order.

Many methods exist to integrate the Schrödinger Equation \cite{WALKER1994,PAGANIN2021,STRANG1968,KOSLOFF1983,DATEO1991,VANDYCK1985}. In~particular, some methods use the Fourier transform explicitly, but~usually as  a trick to make calculations easier, while in others such as split-step, multi-slice, or~Fourier space filtration~\cite{TAHA1984,BANDRAUK1992,HANSSON2016}, the~Fourier transform plays a more fundamental role relating to the Fourier dual~spaces.

In this study, similar to split-step, we break the Hamiltonian into kinetic and potential energy portions and then employ a Recursive Fourier Transform (RFT) technique in a novel way to decouple the second-order integral from the first, bypassing the need to invoke time ordering. Thus, we can represent TDSE, plausibly to any order, as~a convolution equation by invoking the Convolution Theorem. This presentation aims to offer an efficient second-order analytical solution to the TDSE, while also emphasizing the method's utility as a first principle rather than a calculational trick~\cite{NELSONISAACS2021}.

Using this technique to efficiently and precisely calculate the spectral response of a time-limited perturbation has relevance for recent advances in single-photon generation~\mbox{\cite{HONG1987, DAVIS2018,MOSLEY2008,MULLER2017}}, quantum computing~\cite{TAMMALAIBACHER2015, TAMMALAIBACHER2018, TAMMA2021, Tamma2023, wang2018experimental, TRIGGIANI2023, Cui2012, Zhang2024Optimizing, Asavanant2024Multipartite, Bartolucci2023}, optical traps~\cite{Lu2024,NEUMAN2004,Perez-Garcia2023}, quantum cryptography~\cite{Panda2024Quantum}, quantum tunneling and microscopy~\cite{LOUNIS2014,GOTTLIEB2006,Grewal2024,DessaiKulkarni2022,Gaida2024}, quantum information and entanglement~\mbox{\cite{cao2024multiqubit,Kawasaki2024,Kawasaki2024HighRate}}, high harmonic generation~\cite{Nishidome2024,Majidi2023,Farkas1992}, ultra fast light pulse generation~\cite{LEWENSTEIN1994,Ryabikin:2023,HANSCH1990}, and~ measurements of gravity~\cite{ABELE2010,Jenke2011}. Applications to open systems may be possible by extending this method to, for~example, the~Lindblad master Equation \cite{VILLEGAS2016}. This method may also hold pedagogical promise in physics education by expanding the range of calculable use cases for the TDSE~\cite{Krijtenburg2017, Singh2006}.

The procedure will be applied here to basic examples, such as Gaussian potentials, but~it is quite general to any potential whose Fourier transform exists.
One may also find useful application in studying quantum Zeno dynamics,
for instance in suppressing second-order photon exchange relative to first order for the purpose of generating entanglement~\cite{Nodurft2022}.

The procedure may also be useful in quantum error correction techniques by characterizing spectral properties of noise sources leading to the decoherence of~qubits.

A detailed breakdown of the paper is as follows: first-order RFT technique introduction to familiar use cases (Section \ref{hdr:TDSE1stOrder}), a new second-order RFT decoupling technique (Section~\ref{hdr:TDSE2ndOrder}), and experimental and theoretical applications (Section \ref{hdr:discussion}). Appendices with supplemental sections on basic definitions and concepts (Appendices \ref{hdr:FT_def} and \ref{hdr:domainExamples})
, mathematical property examination (Appendix \ref{hdr:funcanalysis}), numerical accuracy of the method (Appendix \ref{hdr:erroranalysis}), and interpretation of the results (Appendix \ref{hdr:interpretation}) are also~included.

\section{Methods}\label{xiaoyi}

The TDSE will now be evaluated to first and second order using the recursive Fourier transform (RFT) method.
We use as a starting point the standard formulation of TDSE, for~instance as in~\cite{SakuraiNapolitano2011}.

The matrix element for the transition between the initial state $\ket{\omega_i}$ and the final state $\ket{\omega_f}$ is
\begin{align}
    \label{eqn:DiracTDSE2}
    \begin{split}
    \cnawtl
    &= \sum_i \Big( \bra{\omega_f}(1-\frac{i}{\hbar} \int_{\TLowerLimit}^{\TUpperLimit} dt_1 \hat{V}_I(t_1) \\
    &\qquad \qquad +(-i/\hbar)^2 \int_{\tLowerLimit}^{\TUpperLimit} dt_1 \int_{\TLowerLimit}^{\tUpperLimit} dt_2 \hat{V}_I(t_1)\hat{V}_I(t_2)+\ldots )\ket{\omega_i}\Big)   \fn,
    \end{split}
\end{align}
where the potential $\hat{V}_I$ is written in the interaction picture.
The potential has both an operator component and a continuous time dependence. We will focus our discussion on the~latter.

\subsection{Formulating the First-Order Time Dependent Schr\"{o}dinger Equation through Recursive Fourier Transforms (RFT)} \label{hdr:TDSE1stOrder}

The first-order term in Equation \eqref{eqn:DiracTDSE2} has the well-known approximation~\cite{Tokmakoff2014} %pg.2-29 (aka 45)
\begin{align}
    \begin{split}
    \label{eqn:FirstOrderFormula}
    \cnawtl &\propto \frac{1}{i \hbar}\sum_i \int_{0}^{\Time} dt_1 \braket{\omega_f|\hat{V}|\omega_i}V(t_1) e^{i (\omega_{f}-\omega_{i}) t_1} \constA \\
    &\sim \sum_i V_{fi} \constA \tilde{V}(\omega_f - \omega_i)
    \end{split}
\end{align}
where the symbol $\sim$ indicates the Fourier transform, $\constA$ is the amplitude of state $\ket{\omega_i}$ at $t=0$, and~$V_{fi} = \braket{\omega_f|\hat{V}|\omega_i}$ is the matrix element of the potential that connects the initial and final states and~will be insignificant to the current discussion. In~the second line, a~standard approximation was made by allowing the time interval to become infinite in both directions, ``asymptotic time,'' so that Equation \eqref{eqn:FirstOrderFormula} becomes the Fourier transform of the potential.
This relationship to the Fourier transform becomes further intriguing when we write Equation \eqref{eqn:FirstOrderFormula} in a suggestive way,
\begin{align}
    \begin{split}
    \label{eqn:FirstOrderA}
    \cnawtl &=  \frac{1}{i\hbar}\int_{\TLowerLimit}^{\TUpperLimit} dt_1 e^{i \omega_f t_1}  V(t_1)  \sum_i e^{-i \omega_i t_1} \braket{\omega_f|\hat{V}|\omega_i} \, \constA .
    \end{split}
\end{align}

Clearly this expression involves a transformation (though not formally a Fourier-type transform) from a frequency representation to a time representation, and~a subsequent transformation back to a frequency~representation.

Based on this intuition, we modify Equation~\eqref{eqn:FirstOrderFormula} to not assume asymptotic time by inserting an indicator function (or mask) that is non-zero only within the specified time range, $0 \rightarrow \Time$,
\begin{align}
    \begin{split}
    \label{eqn:FirstOrderB}
    \cnawtl &= \frac{1}{i\hbar} \sum_i V_{fi} \, \constA   \int_{-\infty}^{\infty} d t_1 e^{i \omega_f t_1} rect\left(\frac{t_1}{\Time}-\frac{1}{2}\right) V(t_1) e^{-i \omega_i t_1}.
    \end{split}
\end{align}

Everything on the right-hand side is written in the time domain over parameter $t_1$, but~by writing 
%$rect(t_1)$, $V(t_1)$, and $\exp{(-i\omega_i t_1)}$ 
each factor as the inverse Fourier transforms of their Fourier transforms,
\begin{align}
    \begin{split}
        rect\left(\frac{t_1}{\Time}-\frac{1}{2}\right) &= \underset{\omega\ra t_1}{\iscrT} \Big\{ \Time \exp{(i \omega \Time/2)} sinc(\omega \Time/2)\Big\} \\ 
        V(t_1) &= \underset{\omega\ra t_1}{\iscrT} \Big\{ \tilde{V}(\omega) \Big\} \\
        e^{-i \omega_i t_1} &= \underset{\omega\ra t_1}{\iscrT} \Big\{ \delta(\omega-\omega_i) \Big\}  ,
    \end{split}
\end{align}
we can write the integral over $t_1$ as follows:
\begin{align}
    \label{eqn:FirstOrderC}
    \underset{t_1 \ra \om}{\scrT} \Bigg\{ \underset{\omega\ra t_1}{\iscrT} \Big\{ \Time \exp{(i \omega \Time/2)} sinc(\omega \Time/2)\Big\} \underset{\omega\ra t_1}{\iscrT} \Big\{ \tilde{V}(\omega) \Big\} \underset{\omega\ra t_1}{\iscrT} \Big\{ \delta(\omega-\omega_i) \Big\}  \Bigg\}\Bigg|_{\om = \omega_f} ,
\end{align}
where $sinc(x) \equiv sin(x)/x$, and~the symbol $\underset{t_1 \ra \om}{\scrT}$ is an obvious notation that makes explicit that the transform is converting from an expression in $t_1$ to an expression in $\om$. 

Note that the outer operation in Equation \eqref{eqn:FirstOrderC} is not technically a Fourier transform to $\om$ but rather a projection onto a single $\exp{(i\omega_f t_1)}$ basis state.
To accomplish this, we computed a Fourier transform to switch to a continuous energy basis and then evaluated the result at a discrete energy $\omega=\omega_f$. Conceptually, this is important because $\omega$ is a dummy convolution variable that is replaced by the measurable energy $\omega_f$.

By applying the convolution theorem to Equation \eqref{eqn:FirstOrderC} and inserting into Equation \eqref{eqn:FirstOrderB}, we obtain an expression for the first-order transition amplitude,
\begin{align}
    \boxed{ \cnawtl =  \frac{ \Time }{2\pi i \hbar}  \sum_i V_{fi} \, \Big( e^{i \om \Time/2} sinc(\om \Time/2) \ast \tilde{V}(\om) \ast \delta{(\om - \omega_i)} \Big)\Big|_{\om = \omega_f} \constA  }.
    \label{eqn:mainResult1}
\end{align}
This is the first main result, the first order spectral response to the time-dependant perturbation $V$. Comparing with Equation \eqref{eqn:FirstOrderFormula}, we observe variations in the frequency domain with a ``spatial frequency'' $4\pi / \Time$, where $\Time$ is the duration of the measurement (see Figure~\ref{fig:Gaussian_Potential}b).

%\subsubsection{Examples}

Two standard cases will be considered to illustrate the validity of the result~above.

\subsubsection{Example: Gaussian-Kicked Harmonic~Oscillator}
%This section is duplicated from "DUPE" in the "Domains" appendix

%REF: eqn. 2-29 p.46 Tokmakoff or 1RabiOscillations
A simple system to consider is the harmonic oscillator ``kicked'' by a small Gaussian pulse,
\begin{align}
    \begin{split}
    \label{eqn:gaussKickA}
    V(t) \propto e^{-\frac{t^2}{2 \tau^2}},
    \end{split}
\end{align}
where $\tau$ is the characteristic time of the Gaussian. For~instance, this can represent a cold atom in an optical trap~\cite{Lu2024,NEUMAN2004}.
%https://en.wikipedia.org/wiki/Optical_tweezers#Electric_dipole_approximation
%https://www.ncbi.nlm.nih.gov/pmc/articles/PMC1523313/#:~:text=In%20an%20optical%20trap%2C%20the,to%20the%20gradient%20trapping%20force.
%https://www.researchgate.net/publication/259786021_Fast_transitionless_expansions_of_Gaussian_anharmonic_traps_for_cold_atoms_Bang-singular-bang_control
%https://pubmed.ncbi.nlm.nih.gov/38653978/

To find the transition amplitude from the $i$ to the $f$ state using Equation \eqref{eqn:FirstOrderFormula}, the~asymptotic time approximation results in the expression
\begin{align}
    \begin{split}
    \label{eqn:gaussKickB}
    c^{(1)}(t \rightarrow \infty) \propto \frac{\tau}{\sqrt{\Omega}} e^{-\om^2 \tau^2/2},
    \end{split}
\end{align}
where $\Omega$ is a normalization constant owing to the fixed natural frequency of the oscillator, and $\om$ parameterizes the frequency response of the Gaussian perturbation.\cite{Tokmakoff2014}

Using instead Equation \eqref{eqn:mainResult1}, the~same transition can be~written as

%WHAT IS THE USE OF T AND \TAU HERE?
\begin{align}
    \begin{split}
    \label{eqn:kickedharmosc1}
    \cnawtl &\approx \frac{1}{i\hbar} V_{fi} \, \constA   \int_{-\infty}^{\infty} d t_1 e^{i \omega_f t_1} rect\left(\frac{t_1}{\Time}-\frac{1}{2}\right) e^{-\frac{t_1^2}{2\tau^2}}  e^{-i \omega_i t_1} \\
    &\approx \frac{\tau \Time }{2\pi i \hbar} V_{fi} \, \Big( e^{i \om \Time/2} sinc(\om \Time/2) \ast e^{-\om^2 \tau^2/2} \ast \delta(\om-\omega_i) \Big)\Big|_{\om = \omega_f}  ,
    \end{split}
\end{align}
where $\Time$ is the measurement interval, and~$\tau$ is the characteristic width of the Gaussian.
Equation \eqref{eqn:kickedharmosc1} is valid under the usual conditions necessary for a Taylor series (weak interaction), but unlike
Equation~(\ref{eqn:gaussKickB}), it does not make the asymptotic time~approximation.

\begin{figure}[H]
  
     \begin{subfigure}[b]{0.45\textwidth}
     
            		\includegraphics[width=\textwidth]{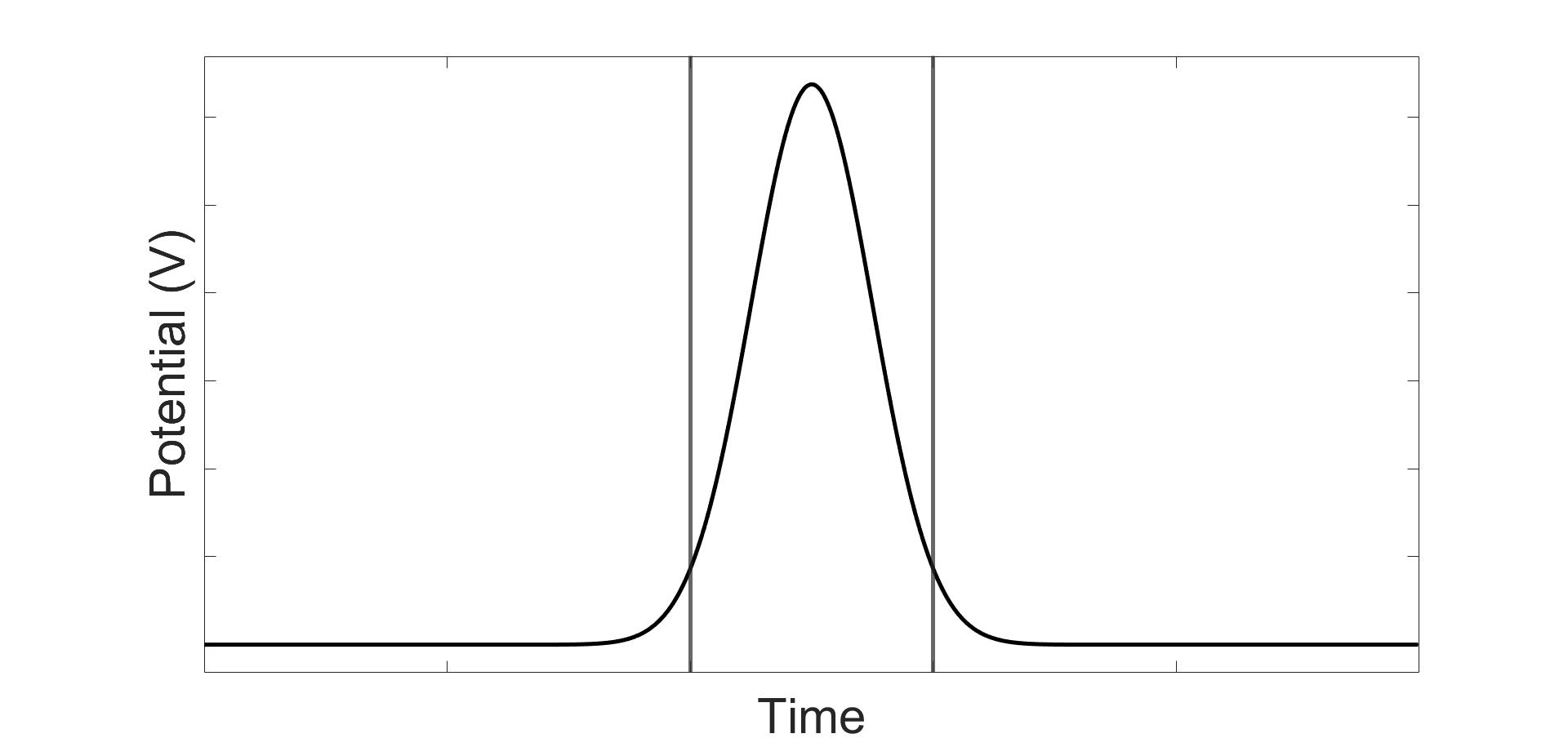} 
            		\caption{ } 
    \end{subfigure}  \hspace{6pt}  
    \begin{subfigure}[b]{0.45\textwidth}       
		\includegraphics[width=\textwidth]{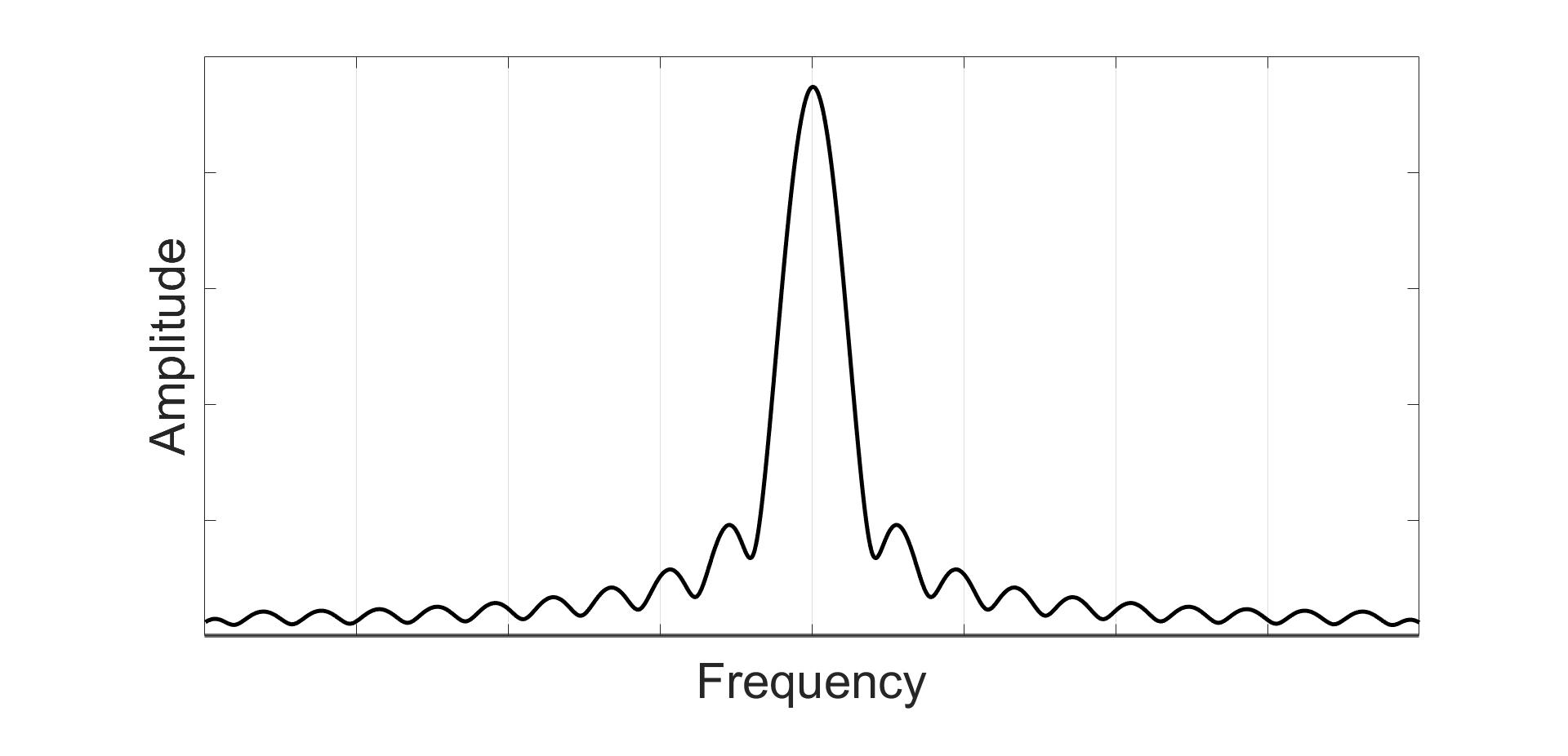}
		\caption{ }
		 \label{fig:mod-sinc}    
    \end{subfigure}\vspace{6pt}
        \caption{In the first order TDSE in Eq. \ref{eqn:mainResult1}, a Gaussian potential's tails are truncated by the measurement. (a) The integration windows for the TDSE are marked as vertical lines. The tails of the Gaussian are excluded, leading to ringing in the frequency domain (not shown). (b) A Gaussian potential in $\om$-space convolved with $sinc(\om \Time /2)$ in Eq. $\ref{eqn:kickedharmosc1}$ shows the variations in the frequency domain with ``spatial frequency'' $4\pi / \Time$.
        %Energy scale is in units f $\omega_0 = 4\pi/\tau$
        } \label{fig:Gaussian_Potential}
    \end{figure}

The effect of convolution is to add a small ripple to the Gaussian (see Figure~\ref{fig:Gaussian_Potential}b). This ripple was ignored in the standard approach (Equation \eqref{eqn:gaussKickB}), when the limits of integration are arbitrarily set to~infinity.

\subsubsection{Example: Fermi's Golden~Rule} \label{hdr:FGR1}
It will next be verified that Equation \eqref{eqn:mainResult1} reduces to the well-known Fermi Golden Rule. A~typical example involves calculating the transition probability for
an electron around the stationary atom absorbing a photon and transitioning from a bound state to a continuum of states. The~Hamiltonian is
\begin{align}
    \label{eqn:fermiRule1}
    \begin{split}
    \hat{H} &= \hat{H}_0  \text{ for } t \leq 0 \\
    \hat{H} &= \hat{H}_0 + \hat{V}(t) \text{ for } t > 0\\
    \hat{V}(t) &= 2 \hat{V}_0 cos(\omega_d t)
    \end{split}
\end{align}
where $\hat{V}(t)$ is the time-dependent perturbation, and $\hat{H}_0$ and $\hat{V}_0$ are independent of~time.

The standard integral in Equation \eqref{eqn:FirstOrderFormula} results in the following $i\rightarrow f$ transition amplitude:
\begin{align}
    \label{eqn:FermiPerturbCoeffOld}
    \begin{split}
    %c_f^{(1)}(T)
    \cnawtl&= \frac{1}{i\hbar} \int_{0}^{T} dt_1 e^{i (\omega_f-\omega_i) t_1} (e^{i \omega_d t_1}+e^{-i \omega_d t_1})  \braket{\omega_f|\hat{V}|\omega_i}\\
    &= \frac{V_{fi}}{i\hbar} \left( \frac{e^{i(\omega_{fi} + \omega_d)T} - 1}{i(\omega_{fi} + \omega_d)} + \frac{e^{i(\omega_{fi} - \omega_d)T} - 1}{i(\omega_{fi} - \omega_d)}\right)  \\
    &\approx \frac{V_{fi}}{i\hbar} e^{i\frac{(\omega_{fi}-\omega_d)}{2}T}
     \left(\frac{sin((\omega_{fi} - \omega_d)T/2)}{(\omega_{fi} - \omega_d)}\right),
    \end{split}
\end{align}
where $\omega_{fi}\equiv \omega_f-\omega_i$. We dropped the first term, as~is customary, in~favor of the second term, which dominates around the resonant frequency $\omega_{fi} \approx \omega_d$ \cite{ZHANG2016}.

Equation~\eqref{eqn:mainResult1} obtains the same result. Computing 
\begin{align*}
    \scrT \{V(t)\} \approx \frac{V_0}{2}\left(\delta(\om-\omega_d) + \delta(\om+\omega_d)\right),
\end{align*}
\textls[-15]{and dropping the $+$ term for the same reason as above, gives}
\begin{align}
    \begin{split}
    \label{eqn:FermiPerturbCoeffNew}
    \cnawtl &= \frac{1}{2\pi i\hbar} V_{fi} \, \Big( \Time e^{i \om \Time/2} sinc(\om \Time/2) \ast \delta(\om-\omega_d) \ast \delta{(\om - \omega_i)} \Big)\Big|_{\om = \omega_f} \\
    &= \frac{1}{2\pi i\hbar} V_{fi} \, \Time e^{i (\omega_f - \omega_i - \omega_d) \Time/2} sinc((\omega_f - \omega_i - \omega_d) \Time/2) 
    \end{split}
\end{align}
which is the standard result (up to constant factors).

%%%%%%%%%%%%%%%%%%%%%%%%%%%%%%%%%%%%%%%%%%

\subsection{Decoupling the Second-Order TDSE through~RFT}\label{hdr:TDSE2ndOrder}

Having examined the familiar first-order result using recursive Fourier transform methods,
we now derive our second main result: an expression for the second-order term in the TDSE~expansion.

The integrals for the second-order amplitudes are more complicated because the upper limit of integration for the nested integral is the integration parameter for the outer integral $t_1$. 
Starting from Equation \eqref{eqn:DiracTDSE2}, by~inserting a discrete basis of equally spaced states, the~second-order transition amplitude is
\begin{align}
    \label{eqn:direct-integration}
    \begin{split}
    \cnats =&\frac{1}{ (i\hbar)^2}\sum_{ki} \int_{\TLowerLimit}^{\TUpperLimit} dt_1 \int_{\tLowerLimit}^{\tUpperLimit} dt_2 
    \bra{\omega_f}\hat{V}_I(t_1)\ket{\omega_k}\bra{\omega_k} \hat{V}_I(t_2)\ket{\omega_i} \constA 
    \\
    =&\frac{1}{ (i\hbar)^2} \int_{\TLowerLimit}^{\TUpperLimit} dt_1 \Big\{ e^{i \omega_f t_1}  V(t_1)\sum_k e^{-i \omega_k t_1} V_{fk} \\ &\qquad \qquad \qquad \int_{\tLowerLimit}^{\tUpperLimit} dt_2 e^{i \omega_k t_2}V(t_2) \sum_i e^{-i \omega_i t_2} V_{ki} \, \constA \Big\}
    \end{split}
\end{align}

The integrals are therefore coupled, and~the method in Section~\ref{hdr:TDSE1stOrder} must be modified. This is a Dyson series and~was decoupled by Dyson by introducing the time-ordering operator. This is used widely in quantum field theory~\cite{DYSON1952}.

Here, the integrals will be decoupled in a new way in the following four~steps.
We assume that the spectra of the energy eigenstates $\omega_k$ are discrete. For~simplicity, we consider only the case in which they are equally spaced, that is, a~simple harmonic oscillator. Then, we can write 
\begin{align} 
    \label{eq:omk}
    \omega_{k} = k \groundom, 
\end{align}
for integers $k$. 

\begin{myStep}
    \label{step:1}
    Apply the convolution theorem to the nested integral
\end{myStep}
The limits of integration of the nested integral are extended to infinity, using a rectangular mask, as~in Equation \eqref{eqn:FirstOrderB},
\begin{align}
    \label{eqn:2ndOrder_step3}
    \begin{split}
    \cnats=& \frac{1}{ (i\hbar)^2}\int_{\TLowerLimit}^{\TUpperLimit} d t_1 e^{i \omega_f t_1} V( t_1) \sum_{k} e^{-i \omega_k t_1}  V_{fk} \\
    &\,\Big \{\int_{-\infty}^{\infty} d t_2 e^{i \omega_k t_2} rect\left(\frac{t_2}{ \DeltaT_1}-\frac{1}{2}\right) V(t_2) \sum_i e^{-i \omega_i t_2}  V_{ki} \, \constA \Big\}
\end{split}
\end{align}

Because we truncated the signal using a $rect(t)$ mask, this step was exact. The~integrals are still coupled via $\DeltaT_1$, but the coupling now parameterizes the width of the mask rather than the integration domain.

By following the steps in Equation \eqref{eqn:FirstOrderB}, we can write each factor in the integrand of the second line of Equation \eqref{eqn:2ndOrder_step3} in the frequency domain,
\vspace{-10pt}
\begin{adjustwidth}{-\extralength}{0cm}
\centering %% If there is a figure in wide page, please release command \centering
\begin{align}
    \label{eqn:2ndOrder_step4}
    \begin{split}
    \cnats =& \frac{1}{2\pi (i\hbar)^2}\int_{\TLowerLimit}^{\TUpperLimit} d t_1 e^{i \omega_f t_1} V( t_1) \sum_{ki} e^{-i \omega_k t_1}  V_{fk}V_{ki} \constA \\
    & \underset{t_2\ra \omega_k}{\scrT}\Big\{ \underset{\om \ra t_2}{\iscrT}\{ \DeltaT_1 \exp{(i \om \DeltaT_1/2)} sinc(\om \DeltaT_1/2) \cdot \underset{\om \ra t_2}{\iscrT} \{ \tilde{V}(\om)\} \underset{\om \ra t_2} {\iscrT} \{\delta(\om-\omega_i)\}\Big \} ,
\end{split}
\end{align}
\end{adjustwidth}
and apply the convolution theorem,
\begin{align}
    \label{eqn:2ndOrder_step5}
    \begin{split}
    \cnats =& \frac{1}{2\pi (i\hbar)^2} \int_{\TLowerLimit}^{\TUpperLimit} d t_1 e^{i \omega_f t_1} V( t_1) \sum_{ki} e^{-i \omega_k t_1}  V_{fk}V_{ki} \constA \\
    & \qquad \left( \DeltaT_1 e^{i (\om-\omega_i) \DeltaT_1/2} sinc((\om-\omega_i) \DeltaT_1/2) \ast  \tilde{V}(\om) \right) \Big|_{\om=\omega_k}
\end{split}
\end{align}

In evaluating the Fourier transforms, we have transformed bases from the original parameter of integration, $t_2$, to~$\om$ and then to $\omega_k$, an~intermediate basis of energy states. Note that the expression inside the parenthesis on the last line of Equation \eqref{eqn:2ndOrder_step5} is a continuous distribution in a dummy parameter $\om$, evaluated at a specific value $\om=\omega_k$ after performing the convolution.

The nested integral is now a convolution in $\om$-space, but~the $sinc$ function's width depends on $t_1$, which is coupled to the outer integral. How do we compute a convolution of a signal whose shape is changing as $t_1$ is  integrated over?

\begin{myStep}
    \label{step:2}
    Discretize the integral over $\DeltaT_1$ as a Riemann sum and move it inside the sum over $k$ and $i$
\end{myStep}
It is easier to handle Equation \eqref{eqn:2ndOrder_step5} by writing the integral over $\DeltaT_1$ as a Riemann sum of step size $\Delta T$, and~rearranging the sums (switching the order of the sum and the integral in an infinite series can have unpredictable effects on the convergence of the series in general, but~for our purposes we only examine the second-order expansion; this poses the same limitation on validity as other variational approaches such as Feynman diagrams): 
\vspace{-10pt}
\begin{adjustwidth}{-\extralength}{0cm}
%\centering %% If there is a figure in wide page, please release command \centering
\begin{align}
    \label{eqn:2ndOrder_step6}
    \begin{split}
    \cnats =& \frac{1}{2\pi (i\hbar)^2}\sum_{ki} V_{fk}V_{ki}\constA  \sum_{n=n_i}^{n_f}
    \Delta T \, e^{i \omega_f n \Delta T} V(n \Delta T) e^{-i \omega_k n \Delta T} \\
    & \qquad \Big( n\Delta T \exp{(i(\om-\omega_i) n\Delta T/2)} sinc((\om-\omega_i) n\Delta T/2) \ast  \tilde{V}(\om) \Big) \Big|_{\om=\omega_k}
    \end{split}
\end{align}
\end{adjustwidth}

Because the second line is a distribution in the frequency domain evaluated at a specific point, it is simply a c-number for each term in the Riemann~sum.

\begin{myStep}
    \label{step:3}
    Allow the variation over time to vary the width of the distribution $sinc(\om n\Delta T)$
\end{myStep}
Here is the central insight to decouple the integrals. 
For each step in the Riemann sum over $n$ (coupling variable), we identify the expression in the second row of Equation \eqref{eqn:2ndOrder_step6}, in~the continuous limit $n\Delta T \rightarrow \DeltaT_1$, as~having the form of an ``impulse response'' (in the time domain),
%$k_n = h(nT)$
\begin{align}
    \label{eqn:impulseResponse}
    \begin{split}
    \boxed{ \psib[\DeltaT_1] \equiv \Big(  \exp{(i(\om-\omega_i) \DeltaT_1/2)} \frac{sin((\om-\omega_i) \DeltaT_1/2)}{(\om-\omega_i)/2} \ast  \potV \Big) \Big|_{\om=\omega_k} }
    \end{split}
\end{align}

Equation \eqref{eqn:impulseResponse} is the impulse response of the system to a perturbation of duration $\DeltaT_1$, the nested integration variable. The intermediate frequency $\omega_k$ is defined in Eqn. \ref{eq:omk}.

\begin{myStep}
    \label{step:4}
    Apply the convolution theorem to the outer integral
\end{myStep}
Now, we can change the Riemann sum back to an integral over $\DeltaT_1$. Crucially, $\psib$, which is an explicit distribution in $\om$-space, appears inside an integral over time $\DeltaT_1$. We can therefore interpret it as a function of time rather than frequency. 
We can now repeat the earlier technique of extending the integration domain in the first-order to $\pm \infty$ and inserting a rectangular function of width $\Time$,
\begin{align}
\label{eqn:2ndOrder_step8}
    \begin{split}
    \cnats =& \frac{1}{2\pi (i\hbar)^2} \sum_{ki} V_{fk}V_{ki}\constA  \int_{-\infty}^{\infty}
    d t_1 e^{i \omega_f \DeltaT_1} rect\left(\frac{t_1}{ \Time}-\frac{1}{2}\right) V(t_1)e^{-i \omega_k \DeltaT_1} \psib[t_1].
    \end{split}
\end{align}

\begin{figure}[H]	
	 
		\hspace{-4pt}\includegraphics[width=\linewidth]{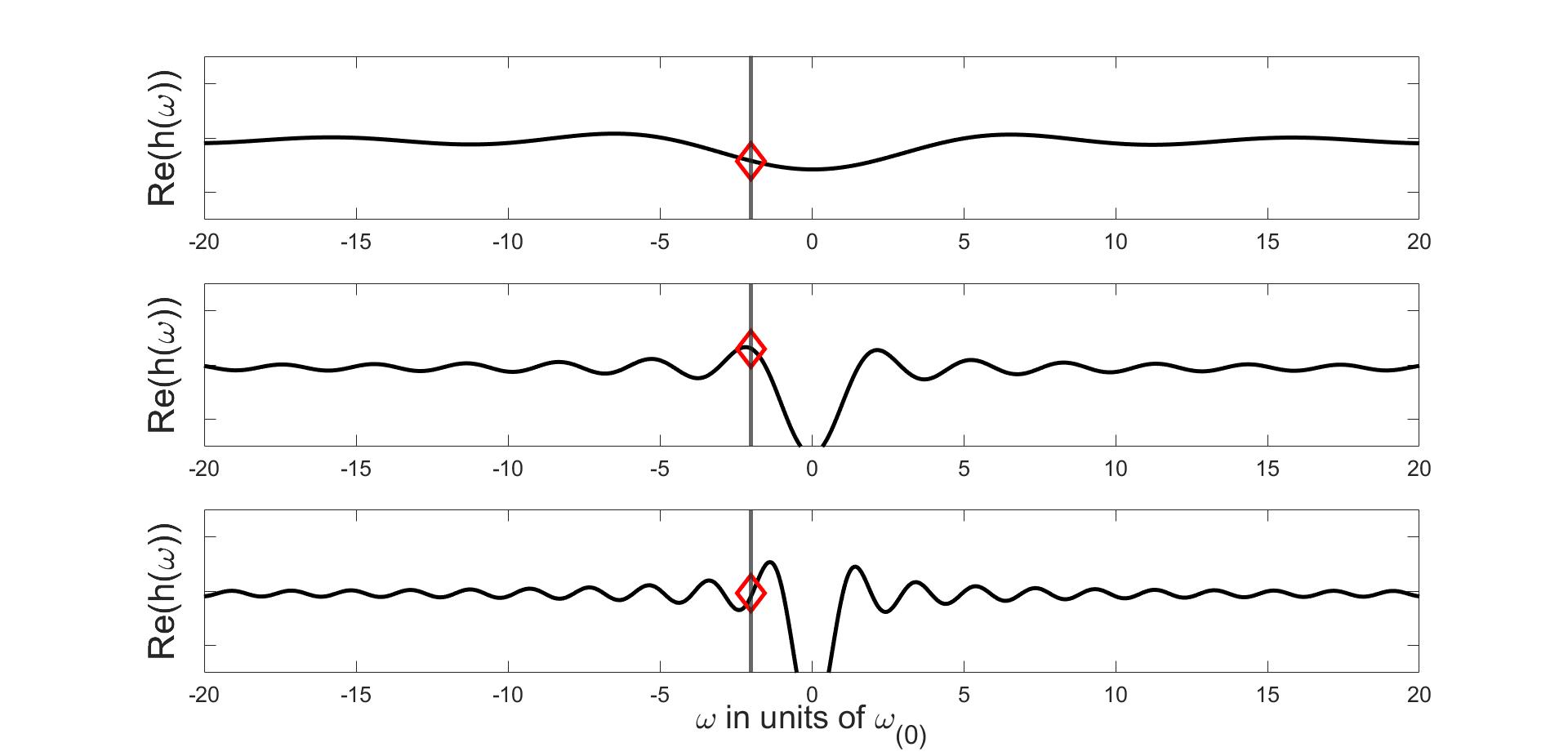}
		\caption{The width of the $sinc$ function in the impulse response depends on $\DeltaT_1$. As $\DeltaT_1$ increases from top to bottom in the figures, we take samples at $\omega=\omega_k$ (defined in Eqn. \ref{eq:omk}, shown as a diamond on vertical line in the figure; here $k=-2$). The~sample values trace out another $sinc$ function.} \label{fig:sinc1}
\end{figure}

The effect of the outer integral over $t_1$ on the nested integral is to vary the width $t_1$ of the impulse response across the duration of the measurement window from $\tLowerLimit \rightarrow \Time$, and~sample it at $\omega_k$ to generate $\psib[\DeltaT_1]$ (see Figure~\ref{fig:sinc1}).

\begin{figure}[H]	
	 
		\hspace{-2pt}\includegraphics[width=0.95\linewidth]{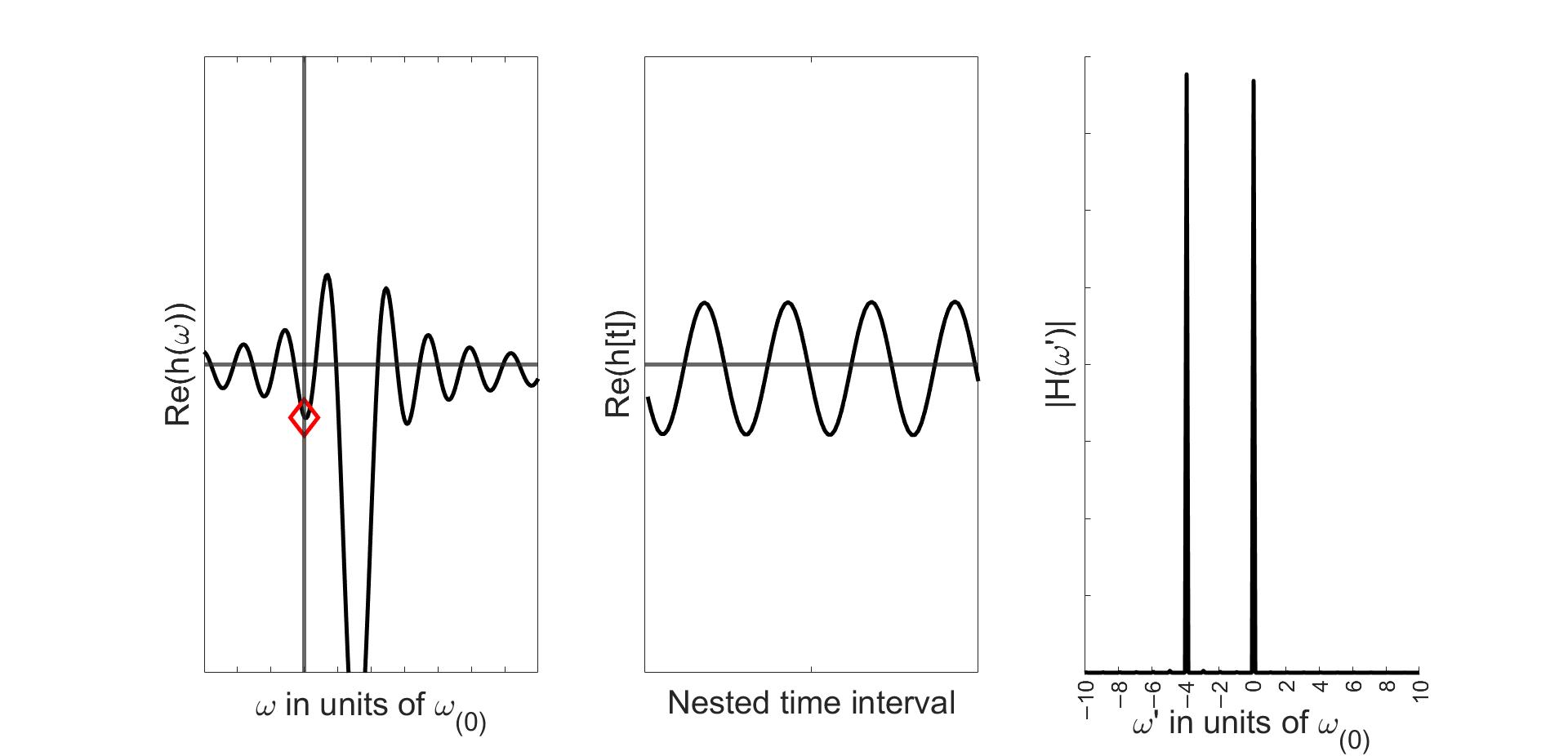}
		\caption{We interpret Equation \eqref{eqn:impulseResponse} as a function of $\DeltaT_1$ instead of $\omega$. (\textbf{Left}) Nested $sinc$ distribution (second line in Equation \eqref{eqn:2ndOrder_step5}), varied in width over $\DeltaT_1$, while repeatedly sampled at $\omega_k=-4\groundom$ (vertical line with diamond marking the sample value). As~the width of the $sinc$ decreases, its height grows linearly with $\DeltaT_1$, so the height of the sample oscillations is constant. (\textbf{Middle}) The samples ($\psib[\DeltaT_1]$) oscillate as $\DeltaT_1$ is varied. (\textbf{Right})~The Fourier transform $\PSIB(\omega')=\scrT\{\psib[\DeltaT_1]\}$ is a series of spikes representing second-order impulses. These $\delta$-functions are convolved in Equation \eqref{eqn:mainResult2} to place a copy of the outer $sinc$ at each~impulse. Note that for illustrative purposes the potential was ignored, i.e. chosen such that $\tilde{V}=\delta(\omega)$.} \label{fig:mainresult2}
\end{figure}

Because of Step \ref{step:3}, everything inside the $t_1$ integral in Equation \eqref{eqn:2ndOrder_step8} can be treated as a distribution in $t_1$, and~the convolution theorem can be used again. 
The $t_1$ integral becomes a Fourier transform by explicitly writing each factor in the $\om$-domain,
\vspace{-10pt}
\begin{adjustwidth}{-\extralength}{0cm}
%\centering %% If there is a figure in wide page, please release command \centering
\begin{align}
    \begin{split}
    \label{eqn:2ndOrder_step9}
    \underset{t_1\ra \omega_f}{\scrT}\Bigg\{ 
    \underset{\om \ra t_1}{\iscrT} \Big\{ \Time e^{i \om \Time/2} sinc(\om \Time/2)\Big\} \cdot \underset{\om \ra t_1}{\iscrT} \Big\{\tilde{V}(\om)\Big\}\underset{\om \ra t_1}{\iscrT} \Big\{ \delta(\om - \omega_k)\Big\} \cdot \underset{\om \ra t_1}{\iscrT}\Big\{\PSIB(\om)\Big\}    \Bigg\},
    \end{split}
\end{align}
\end{adjustwidth}
where $\PSIB(\omega) \equiv \underset{t_1\ra \omega}{\scrT} \{ \psib[t_1] \}$ is the Fourier transform of the impulse response, also called the amplitude transfer~function (see Figure~\ref{fig:mainresult2}).

The final $\om$-domain expression for second-order transition amplitude from $i \rightarrow f$ is
\begin{align}
    \begin{split}
    \label{eqn:mainResult2}
    \boxed{ \cnats = \sum_{ki} \constB \Bigg\{\Time  e^{i \omega_f \Time/2} sinc\left(\frac{\omega_f \Time}{2}\right) \ast \tilde{V}(\omega_f) \ast \delta(\omega_f-\omega_k) \ast \PSIB(\omega_f) \Bigg\} }
    \end{split}
\end{align}
where
\begin{align}
    \begin{split}
    \label{eqn:mainResult3}
    \boxed{ \PSIB(\om) = \underset{\DeltaT_1 \ra \om}{\scrT} \Big\{
    \Big(  \exp{(i(\om'-\omega_i) \DeltaT_1/2)} \frac{sin((\om'-\omega_i) \DeltaT_1/2)}{(\om'-\omega_i)/2} \ast  \tilde{V}(\om') \Big) \Big|_{\om'=\omega_k}
    \Big\} }
    \end{split}
\end{align}

For notational simplicity, we have defined $\constB \equiv \constA V_{fk}V_{ki}/(2\pi i\hbar)^2$.

This is the second main result, expressing the second-order contribution to the transition amplitude owing to an arbitrary time-limited perturbation of a system with evenly spaced energy eigenkets $\ket{\om_k}$.

A detailed analysis of Eqs. \ref{eqn:mainResult2} and \ref{eqn:mainResult3} has been placed in App. \ref{hdr:funcanalysis}.

\begin{figure}[H]	
	 
		\includegraphics[width=0.95\linewidth]{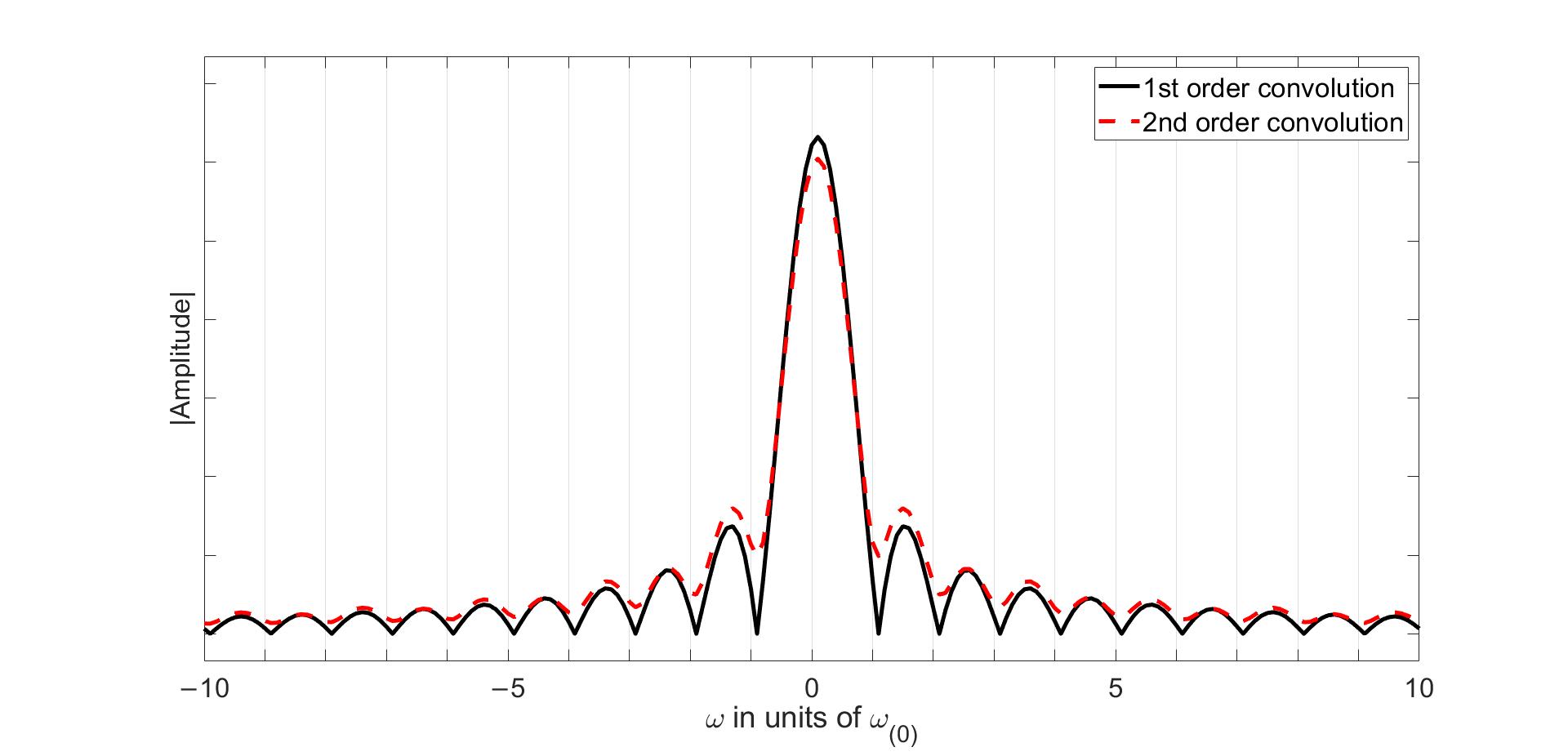}
		\caption{Comparison of first and second-order transition amplitude, relative to initial state $\omega_i$, calculated with the RFT method introduced here. The~second-order calculation computes paths through intermediate energies at $\pm20\groundom$. For~the second-order calculation, the~central peak is reduced, the~wings are amplified, and~the minima are increased. Only the range $\pm10\groundom$ is shown, but~the contributions from terms outside of this range have a significant effect on the accuracy of the result. Not drawn to scale: the second-order contribution is in reality reduced by a factor of $\hbar$ relative to~first-order.} \label{fig:ordercompare1}
\end{figure}
\unskip

\subsubsection{Example: Second-Order Harmonic Perturbation Golden~Rule} \label{hdr:FGR2}
%Michael Fowler, pg 7, top equation, alo pg 8. Page 7 is the 1st order calc which shows how to get te delta function, and page 8 is the 2nd order calc which shows how to apply at second-order.
%time-dependent perturbation theory
To illustrate the results of Section~\ref{hdr:TDSE2ndOrder}, we compare the recursive Fourier transform method with the standard second-order approach~\cite{baym1969,Fowler2007}.

Consider the ramped up oscillating potential,
\begin{align}
    \label{eqn:rampedOscV}
    V(t) = e^{\epsilon t} e^{-i \omega_d t},
\end{align}
where $\epsilon$ is a small positive constant which ensures ramp up of the potential from $t = -\infty$, and~$\omega_d$ is the driving frequency of the potential. This can model, for~instance, light interacting with an atom~\cite{SakuraiNapolitano2011}.

Via the traditional application of the TDSE at second-order, we integrate the potential twice (Equation \eqref{eqn:DiracTDSE2}) to obtain
\begin{align}
    \label{eqn:trad2ndOrderAmp}
    \cnats &= \left(\frac{1}{i\hbar}\right)^2 \frac{e^{i(\omega_{f}-\omega_{i}-2\omega_d)\Time }e^{2 \epsilon \Time}}{\omega_{f}-\omega_{i}-2\omega_d-2i\epsilon } \sum_k \frac{V_{fk}V_{ki}}{\omega_k - \omega_i -\omega_d - i\epsilon}.
\end{align}

This can be interpreted as an amplitude for a particular transition through an intermediate state $k$, summed over all such possible paths (see \cite{baym1969}).
Taking the time-derivative of the squared amplitude in the small $\epsilon$ limit results in an expression for the transition rate,
\begin{align}
    \begin{split}
    \label{eqn:standard2ndGoldenRule}
    \lim_{\epsilon \to 0}\frac{d}{d\Time}\Big|\cnats\Big|^2 &=\left(\frac{1}{i\hbar}\right)^2  \Big| \sum_k \frac{V_{fk}V_{ki}}{\omega_k - \omega_i -\omega_d - i\epsilon}\Big|^2\delta(\omega_f - \omega_i - 2\omega_d) 
    \end{split}
\end{align}
where the $\delta$-function comes from the small $\epsilon$ limit of 
%Michael Fowler, pg 7, top equation, time-dependent perturbation theory
\begin{align}
    \label{eqn:deltaDef1}
    \lim_{\epsilon \to 0} \frac{2\epsilon}{(\omega_{f} - \omega_{i}- 2\omega_d)^2 + \epsilon^2}
    = \delta(\omega_{f} - \omega_{i}-2\omega_d).
\end{align}

This is the second-order version of Fermi's Golden~Rule.

A similar expression can be achieved using recursive Fourier transforms. Starting with Equations~(\ref{eqn:mainResult2}) and \eqref{eqn:mainResult3}, the~second order transition amplitude is
\vspace{-11pt}
\begin{adjustwidth}{-\extralength}{0cm}
%\centering %% If there is a figure in wide page, please release command \centering
\begin{align}
    \begin{split}
    \label{eqn:rampedOscAmp}
    \cnats &= \left(\frac{1}{i\hbar}\right)^2 \sum_k V_{fk}V_{ki} t e^{i\omega_f t/2}sinc(\omega_f t/2) \ast \tilde{V}(\omega_f) \ast \delta(\omega_f - \omega_k) \ast \PSIB(\omega_f),
    \end{split}
\end{align}
\end{adjustwidth}
where
\begin{align}
    \begin{split}
    \label{eqn:rampedOscTransfer}
    \PSIB(\omega_f) = \underset{t_1 \ra \omega_f}{\scrT}\Big\{ \frac{ e^{i(\om'-\omega_i)\frac{t_1}{2}} sin((\om'-\omega_i)\frac{t_1}{2})}{(\om'-\omega_i)/2} \ast \tilde{V}(\om')\Big|_{\om' \ra \omega_k} \Big\}.
    \end{split}
\end{align}

In Equation \eqref{eqn:rampedOscAmp}, the convolution is over $\omega_f$, while in Equation \eqref{eqn:rampedOscTransfer}, the convolution is over $\omega'$.
For $\epsilon > 0$,  the~Fourier transform of the potential (Equation \eqref{eqn:rampedOscV}) is
\begin{align}
  \begin{split}
    \label{eqn:potentialGoldenRule2}
      \hat{V}(\omega') &= \underset{t \ra \omega'}{\scrT}\{  e^{\epsilon |t|} e^{-i \omega_d t} \} \\
     &=\frac{2\epsilon}{ (\omega' -\omega_{d})^2 + \epsilon^2},
  \end{split}
 \end{align}
 which, in~the limit $\epsilon \rightarrow 0$, becomes $\tilde{V}(\omega') = \delta(\omega' -\omega_{d})$ (see Equation \eqref{eqn:deltaDef1}).
Then, the transfer function in Equations~(\ref{eqn:H_w1}) and \eqref{eqn:H_w2} becomes
\begin{align}
    \begin{split}
    \label{eqn:transferFnGoldenRule2}
    \PSIB(\omega_f) \propto \frac{2 \pi i}{\omega_k - \omega_i - \omega_d} \left( \delta(\omega_f + (\omega_k - \omega_i - \omega_d)) - \delta(\omega_f - \omega_d)\right).
    \end{split}
\end{align}

This creates a series of descending harmonic spikes as in Figure~\ref{fig:Hw_impulses_2}, but~in this case centered on $\omega_d$. This is precisely the content of the summation terms in Equation \eqref{eqn:trad2ndOrderAmp}, but~the new calculation exposes the hidden structure in the harmonics, as~described in Appendix~\ref{hdr:funcanalysis}.

Inserting Equations  \eqref{eqn:potentialGoldenRule2} and  \eqref{eqn:transferFnGoldenRule2} into Equation \eqref{eqn:rampedOscAmp} obtains
\begin{align}
    \begin{split}
    \label{eqn:2ndOrderExample}
    \cnats &= \left(\frac{1}{i\hbar}\right)^2 \sum_k V_{fk}V_{ki} t e^{i\omega_f t/2} sinc(\omega_f t/2) \ast \delta(\omega_f-\omega_d) \ast \delta(\omega_f - \omega_k) \ast  \\
    &\qquad \qquad \frac{2\pi i}{\omega_k - \omega_i - \omega_d} \left( \delta(\omega_f + (\omega_k - \omega_i - \omega_d) - \delta(\omega_f  - \omega_d))  \right) \\
    &= \left(\frac{1}{i\hbar}\right)^2 \sum_k \frac{ 2 \pi i t \, V_{fk}V_{ki}}{\omega_k - \omega_i - \omega_d} \Big( e^{i(\omega_f  - \omega_i -2\omega_d)t/2} sinc((\omega_f  - \omega_i -2\omega_d) t/2) \\
    & \qquad\qquad \qquad\qquad \qquad -  e^{i(\omega_f  - \omega_k -2\omega_d)t/2} sinc((\omega_f  - \omega_k -2\omega_d) t/2) \Big)  
    \end{split}
\end{align}

In the limit of large $t$, the~$sinc$ function and its pre-factor approach a delta function, $\delta(\omega_f - \omega_i -2\omega_d)$.
In the limit of small $\epsilon$, the~factor $\exp(2\epsilon t) \rightarrow 1$. 

The time dependence of Equation \eqref{eqn:2ndOrderExample} is the same as in Equation \eqref{eqn:standard2ndGoldenRule}, only made more explicit by expressing the result  in terms of frequency rather than in terms of energy. The~time derivative (of the amplitude, in~this case) is constant, as~in Equation \eqref{eqn:standard2ndGoldenRule}.

Thus, the new method given in Equation \eqref{eqn:rampedOscAmp} and the traditional method given in Equation \eqref{eqn:standard2ndGoldenRule} give similar results for the measurable transition probability under the given~conditions.

%\section{Results}
\section{Discussion} \label{hdr:discussion}

The recursive Fourier transform method for decoupling the Dyson series has application in both experiment and theory. A~few possibilities are discussed.
%We begin with the theoretical and then cover experimental benefits.

\subsection{Bosonic Sampling and Quantum~Computation}
Bosonic sampling~\cite{TAMMALAIBACHER2015} with indistinguishable photons represents a computational challenge that can only be tackled by quantum computers and thus would demonstrate so-called quantum~supremacy. 

Tamma and Laibacher explain that ``for a given interferometric network, the~interference of all the possible multi-photon detection amplitudes\ldots depends only on the pairwise overlap of the spectral distributions of the single photons'' \cite{TAMMA2016BOSON}.
They emphasize extracting quantum information from the ``spectral correlation landscapes'' of photons~\cite{TAMMALAIBACHER2018}.
So characterizing the frequency spectrum of a single photon is an essential~task.

Various physical properties are related to the spectra of the photon. Further elaborating, Tamma and Laibacher assert that their results reveal the ``ability to zoom into the structure of the frequency correlations to directly probe the spectral properties of the full N-photon input state\ldots'' \cite{TAMMALAIBACHER2018}, where ``single-photon states''
\begin{align}
    \begin{split}
    \label{eqn:singlephoton}
    \ket{\psi} := \int_0^\infty d\om \, \cn(\om) e^{i\om \Delta t_0} \hat{a}^\dagger(\om) \ket{0}
    \end{split}
\end{align}
are characterized by a spectral distribution $\cn(\om)$~\cite{TAMMALAIBACHER2018}.

The indistinguishability of photon pairs, time delays between photons, generation of ultra short photons, and probability of detection in a multi photon experiment can all be related to the spectral~distribution.

The recursive Fourier transform approach we explored in this study allows us to calculate the spectral distributions $\cn(\om)$ of photons with greater precision and efficiency, potentially leading to improvements in the above areas of~research.

\subsection{Quantum Field~Theory}
Dyson decoupled the nested integrals in higher-order TDSE (Equation \eqref{eqn:DiracTDSE2}) by introducing a time-ordering operator that places all operators in order of increasing time from right to left. Then, TDSE can be written as a complex exponential,
\vspace{6pt}
\begin{align}
    \label{eqn:action-integral-1}
    \begin{split}
    U_{fi} &= 1 - i\int_i^f dt V_I(t) + \frac{(-i)^2}{2} \int_i^f dt_1 \int_i^{t_1} dt_2 V_I(t_1)V_I(t_2)+\mathcal{O}(V_I^3)\ldots  \\
    &=  1 - i\int_i^f dt V_I(t) + \frac{(-i)^2}{2} \int_i^f dt_1 \int_i^{f} dt_2 T\{V_I(t_1)V_I(t_2)\} + \mathcal{O}(V_I^3)\ldots  \\
    &= T \Big\{ \exp{\Big[ -i \int_i^f dt V_I(t) \Big]}\Big\}
    \end{split}
\end{align}
where $V_I$ is expressed in the Interaction Picture of Dirac. From~this method, the usual field theory methods for calculating field correlation functions are typically~derived.

In this study, we accomplished decoupling in a novel way, with~no appeal to time ordering. This may be a more efficient method for directly calculating higher-order correlation functions or Feynman amplitudes by using convolution. It also removes the asymptotic time assumption, because~the limited time intervals are computed exactly, without~approximating the integration domain to be~infinite.

%A further point is that perturbation theory applies only for small perturbations, so that the higher order terms in the Taylor expansion of the integrand can be ignored.
If the recursive Fourier transform method allows for efficient calculation of higher order terms, one may be able to relax the constraint for small perturbations and allow for a broader range of potential strengths, moving out of the regime of weak coupling~forces.

\subsection{Bardeen~Tunneling} \label{hdr:bardeen1}
%1/23/23 p. 7-8, also see Gottlieb
%Furthermore, %See 1/23/23 p.13 for math check

Bardeen investigated electron tunneling at a voltage-biased junction between the conductive components. Following \cite{LOUNIS2014}, describe the tunneling potential between tip and ssample of an electron microscope as $V(t)\propto \exp{(\eta t/\hbar)}$, and after making several assumptions (for instance, keep only first-order terms, and small tunneling current), the solution the the TDSE is
\begin{align*}
    c_f(t) &= M_{if} \frac{ e^{-i (\omega_f - \omega_i+ i \eta ) t} }{(\omega_f - \omega_i + i \eta)},
\end{align*}
where $M_{if}$ is the matrix element of the Hamiltonian perturbation. 
The energies $\omega_f$ and $\omega_i$ correspond to the sample and tip of an electron microscope, respectively. Electrons come in with energy $\omega_i$, and~in a junction biased at voltage $V_0$, $\omega_i \rightarrow \omega_i + eV_0$. 

Next one typically estimates the tunneling current as the time rate of change of the tunneling probability, in the limit that $\eta \ra 0$.
\begin{align*}
    I(\omega_f) &= \lim_{\eta\ra 0} \frac{d}{dt}|c_f(t)|^2 \\
        &= |M_{if}|^2 \frac{d}{dt} \frac{ e^{2 \eta t} }{(\omega_f - \omega_i - eV_0)^2 + \eta^2)} \\
        &= 2\pi |M_{if}|^2 \delta(\omega_f - \omega_i - eV_0),
\end{align*}
where we used the identity $\text{lim}(\eta \ra 0) (\eta/(\omega^2 +\eta^2)) = \delta(\omega)$.

Bardeen's formulation is valid under certain conditions~\cite{GOTTLIEB2006}. This result is useful because it describes tunneling in terms of time rather than~space.

A related calculation can be executed using Equation \eqref{eqn:mainResult1} and identifying $\tilde{V}  = \delta(E)$, since the potential is constant in time, and $\hbar \omega_i = E_i+ e V_0$, since the initial state includes and offset due to the bias potential.
\begin{align}
    \begin{split}
    \label{eqn:bardeen1b}
    c_f(t) &= \frac{i\hbar}{2} V_{if} \left(  \frac{ \exp{(i \omega t/2)})\sin( \omega t/2)}{\omega} \ast \delta(\omega - e V_0) \ast \delta(\omega- \omega_i) \right)\Big|_{\omega=\omega_f} \\
    &=  \frac{i\hbar}{2} V_{if} \left(  \frac{ \exp{(i (\omega_f - \omega_i - eV_0) t/2)})\sin((\omega_f - \omega_i - eV_0)t/2)}{\omega_f - \omega_i - eV_0}\right)
    \end{split}
\end{align}

%The potential (in the frequency representation) is identified as $\tilde{V}  = \delta(\omega)$, and~the initial state is $\omega_i+ e V_0$.
With~increasing time, Equation \eqref{eqn:mainResult1} converges to a $\delta$-function.

In the first case, we find that the tunneling current is non-zero when $\omega_f = \omega_i + eV_0$. In the second case, we find that the transition amplitude is non-zero under the same condition, which has the same physical meaning.

The recursive Fourier transform approach efficiently determines the probability amplitude for each outgoing energy mode. This method could permit a more detailed description of the energy transitions across the tunneling barrier, including for arbitrarily short time-scales. For~instance, in~tunneling across a voltage-biased barrier, the~energy profile Equation \eqref{eqn:mainResult1} correlates with the excess kinetic energy profile of an electron ensemble post-barrier-crossing. The~ensemble's velocity profile could be~measured.

\subsubsection*{Example: 2nd Order Bardeen~Tunneling} \label{hdr:bardeen2}
%1/23/23 p. 11-12 Red binder
%In Section~\ref{hdr:bardeen1}, we showed that, to first-order, finding the interaction amplitude via the recursive Fourier transform/convolution approach gives a similar result as Bardeen's approach for the amplitude.
Although a second-order expression is not typically attempted with Bardeen's approach, the~recursive Fourier transform approach to the general TDSE allows us to guess at a second-order result for Bardeen~tunneling.

First, we determine the transfer function for second order Bardeen tunneling. Again using $\tilde{V}  = \delta(E)$ and $\hbar \omega_i = E_i+ e V_0$, 
%using the associations $\omega_i \ra E_i$, $\omega_f \ra E_f$, $\omega_k \ra E_b$.
%Identifying the potential as constant in time,
%$V(t) = e^{ie V_0 t/\hbar}$ (Equation \eqref{eqn:bardeenV}) where $V_0$ is the junction bias, 
%and offsetting the initial energy state by the amount of the voltage bias across the junction, $\omega_i=e V_0/\hbar$, 
Equation \eqref{eqn:mainResult3} obtains
\begin{align}
    \label{eqn:bardeenH}
    \begin{split}
    \PSIB(\omega_f) &= \underset{t_1 \ra \omega_f}{\scrT}\Big\{ \left( \frac{\exp{(i (\om'- e V_0/\hbar) \frac{t_1}{2})} sin((\om'- e V_0/\hbar) \frac{t_1}{2})}{(\om'- e V_0/\hbar)} \ast \delta(\om')\right) \Big|_{\om' = \omega_k}
    \Big\}  \\
    &= \underset{t_1 \ra \omega_f}{\scrT}\Big\{  \frac{\exp{(i(\omega_k - e V_0/\hbar ) \frac{t_1}{2 })} sin((\omega_k - e V_0/\hbar ) \frac{t_1}{2 })}{(\omega_k - e V_0/\hbar)}
    \Big\} 
    \end{split}
\end{align}

Note that $\om'$ is the convolution parameter. As~usual, indices $i,f,k$ correspond to initial, final, and~intermediate energy states, respectively.
Performing the Fourier transform results~in
\begin{align*}
    \begin{split}
    \PSIB(\omega_f) &=  \frac{1}{2\pi i(\omega_k - eV_0/\hbar)}\Big( \delta(\omega_f + (\omega_k-eV_0/\hbar) ) - \delta(\omega_f) \Big).
    \end{split}
\end{align*}

Following the steps in Appendix~\ref{hdr:PSIB}, we write
\begin{align*}
    \delta(\omega_f-\omega_k) \ast \PSIB(\omega_f) = \frac{1}{2\pi i(\omega_k - e V_0/\hbar)} \Big( \delta(\omega_f - e V_0/\hbar) - \delta(\omega_f - \omega_k) \Big)
\end{align*}
so using Equation \eqref{eqn:mainResult2}, the 2nd order Bardeen tunneling amplitude is
%currently \nu = initial and \mu = final state, except not on M_{\mu \nu}. Need to switch
\begin{align}
    \label{eqn:bardeen2b}
    \begin{split}
    \cnats = 2\pi i \Time  \hat{V}_{fk}\hat{V}_{ki} \,
    &\sum_k  \frac{ 1 }{(\omega_k - \frac{eV_0}{\hbar})} e^{i \omega_f \frac{\Time}{2}} sinc\left(\omega_f  \frac{\Time}{2}\right) \\
    &\ast
    \Bigg( \delta(\omega_f - \frac{eV_0}{\hbar}) - \delta(\omega_f - \omega_k) \Bigg),
    \end{split}
\end{align}
where $\hat{V}_{ab}$ denotes the matrix elements of the potential operator, and the convolution variable is $\omega_f$. 
%OLD VERSION: &e^{i (\omega_f  - \frac{eV_0}{\hbar}) 

Equation \eqref{eqn:bardeen2b} incorporates two series of complex $sinc$ functions. The first series is centered at $\omega_f = eV_0/\hbar$. The distributions in the second series are centered at $\omega_f = \omega_k$, descending  in amplitude away from $eV_0/\hbar$.  See Figure~\ref{fig:Hw_impulses_2}.

This represents the distribution of kinetic energies of tunneled electrons described by Equation \eqref{eqn:bardeen2b}.
This result is significant as a transient effect for small times only. It illustrates the usefulness of the RFT method for extending existing methods of calculation.

\subsection{Joint Spectral Amplitude~Function}
%7/19/23 p. 10, 13, 14, 15, 18b,27,32
%8/12/23 p. 3

Recent experiments in quantum optics~\cite{Li2008,CHEN2005,SHARPING2004,GARAYPALMETT2007,KELLER1997,RUBIN1994}
rely on the generation of entangled photons characterized by a joint spectral amplitude function (JSA), $F(\omega_s, \omega_i) $, such that
%Li eqn 5
\begin{align*}
    \ket{\Psi(\omega_s,\omega_i)} \propto \int d \omega_s \int d \omega_i F(\omega_s, \omega_i) \ket{\omega_s} \ket{\omega_i}
\end{align*}

To be concrete, consider 4-wave mixing with signal, idler, and~pump photons given by
%Li eqn 2-4
\begin{align*}
    \begin{split}
    E_{s(i)} (t,z) &= \int d \omega_{s(i)} a_{s(i)}^+ e^{-ik(\omega_{s(i)}) z + i \omega_{s(i)} t} \\
%    E_i (t,z) &= \int d \omega_i a_i^+ e^{-ik_i z + i \omega_i t} \\
    E_p(t,z) &=  e^{-i \gamma P z} \int d \omega_p e^{-\frac{(\omega_p-\omega_{0})^2}{2 \sigma^2}} e^{i k(\omega_p) z - i \omega_p t}
    \end{split}
\end{align*}
%We define $\Delta k \equiv k_{p0} - k_s(\omega_s)- k_i(\omega_i)$.

In this process, two pump photons at frequency $\omega_0$ annihilate to generate two outgoing photons (signal and idler) at frequencies $\omega_0 \pm \Delta$ for some frequency detuning value $\Delta$. This is a statement of energy conservation. The~photons are assumed to be in a non-linear, dispersive medium with wave vector $k(\omega)$ (the non-linearity is contained in the expression with $\gamma$, but~is not important for this derivation).
At first order, the Schr\"{o}dinger equation~gives
\begin{align*}
    \begin{split}
    \ket{\Psi(\omega_s,\omega_i)} & \propto \int_0^T dt \int_{-L}^0 dz E_s  E_i  E_p E_p \ket{0} \\
    &\propto \int d\omega_s d\omega_i \Big(  \int_0^T dt  \int_{-L}^0 dz e^{-i( k_s(\omega_s)z- \omega_s t)}e^{-i( k_i(\omega_i)z- \omega_i t)} E_pE_p \Big)\ket{\omega_s} \ket{\omega_i} \\
    &\propto \int d\omega_s d\omega_i F(\omega_s, \omega_i) \ket{\omega_s} \ket{\omega_i}
    \end{split}
\end{align*}
where $F(\omega_s, \omega_i) $ is the~JSA.

The wave vector mismatch (due to dispersion) is calculated from the Taylor expansion of the wave vectors,
%by defining the detuning parameter, 
%$\Delta \equiv (\omega_s - \omega_i)/2$, so that
\begin{align*}
    \Delta k &= 2k(\omega_p) - k(\omega_s) - k(\omega_i) 
    = (\omega_{p} - \omega_0)^2 k'' - \Delta k(\omega_s,\omega_i)
\end{align*}
where
\begin{align}
    \Delta k(\omega_s,\omega_i) \equiv \frac{k''}{4}(\omega_s - \omega_i)^2.
    \label{eq:waveMismatch}
\end{align}

The JSA can be written as %\cite{KELLER1997}\cite{RUBIN1994}
\begin{align}
    \begin{split}
        F(\omega_s, \omega_i) &=  \int_{-L}^0 dz e^{-i ( k(\omega_s)+ k(\omega_i) ) z}
    \int_0^T dt e^{i(\omega_s+\omega_i)t}
        \\
        &\qquad \Big( e^{-i\gamma P z} \int d \omega_p e^{-\frac{(\omega_p-\omega_{0})^2}{2 \sigma^2}} e^{i k(\omega_p) z - i \omega_p t}\Big)^2 \\
    &=\int_{-L}^0 dz e^{-i (\Delta k(\omega_s, \omega_i)+2\gamma P)  z}
    \int_0^T dt e^{i(\omega_s+\omega_i)t}
        \\
        &\qquad \Big( \int d \omega_p e^{-\frac{(\omega_p-\omega_{0})^2}{2 \sigma^2}(1-2ik'' \sigma^2 z)} e^{ - i \omega_p t}\Big)^2,
    \label{eq:4wave1}
    \end{split}
\end{align}

For Gaussian pump photons, the integral can be evaluated in closed form to obtain an expression (to second order) for the JSA~\cite{Li2008},%\cite{CHEN2005}\cite{SHARPING2004}\cite{GARAYPALMETT2007}
 %See 9/22/23 p.5
\begin{align}
    \begin{split}
    F(\omega_s,\omega_i) &= \int_{-L}^0 dz \frac{1}{\sqrt{1-\frac{ik''}{2}\sigma_p^2 z}} e^{-i(\Delta k +2\gamma P)z} e^{-\frac{(\omega_s+\omega_i-2\omega_{0})^2}{4 \sigma_p^2}} \\
    &\approx e^{i(\Delta k+2\gamma P) \frac{L}{2}} sinc((\Delta k+2\gamma P)\frac{L}{2}) e^{-\frac{(\omega_s+\omega_i-2\omega_{0})^2}{4 \sigma_p^2}},
    \end{split}
    \label{eqn:JSF-LI}
\end{align}
where in the last step, the radical is set to unity using the fiber approximation $|k''|<<\frac{1}{\sigma_p^2 L}$.

Alternately, this expression can be derived using the recursive Fourier transform process developed in this paper. Rewrite the integrals in Equation~(\ref{eq:4wave1}) as
\begin{align*}
    \begin{split}
        F(\omega_s, \omega_i) &=  \int_{-\infty}^{\infty} dz \,rect(\frac{z}{L_0}+\frac{1}{2}) e^{-i \Delta k z}\\
        &\qquad \int_{-\infty}^{\infty} dt \,rect(\frac{t}{T_0}-\frac{1}{2}) e^{i(\omega_s+\omega_i)t}
         \Big( e^{-i \omega_{0} t} e^{-i\gamma P z} \int d \Omega e^{-\frac{\Omega^2}{2 \sigma^2}} e^{i \frac{1}{2}\Omega^2 k'' z - i \Omega t}\Big)^2 
    \end{split}
\end{align*}
where $\Omega \equiv \omega_p-\omega_{0}$.

Using the methods of Section~\ref{xiaoyi}, we write
\begin{align*}
    \begin{split}
        F(\omega_s, \omega_i) =  L_0 T_0         &\int dz e^{-i \Delta k z} \underset{k \ra z}{\iscrT}\{  e^{ik\frac{L_0}{2}} sinc(k\frac{L_0}{2}) \} \underset{k \ra z}{\iscrT}\left\{  \delta( k + 2 \gamma P) \right\} \\
        &\int dt e^{i(\omega_s + \omega_i)t}  \underset{\omega \ra t}{\iscrT}\{ e^{i \omega \frac{T_0}{2}} sinc(\omega \frac{T_0}{2}) \} \underset{\omega \ra t}{\iscrT}\left\{  \delta( \omega - 2 \omega_{0} )  \right\} \\
        &\qquad \left( \int d \Omega \,  e^{-i \Omega t} \exp{(-\frac{\Omega^2}{2 \sigma^2}(1-ik'' \sigma^2 z))} \right)^2
    \end{split}
\end{align*}

Performing the Fourier transform over $z$ first results in a factor $\delta(k  + \Omega^2 k''/2)$, 
and using the convolution theorem, we arrive at
\begin{align}
    \begin{split}
        F&=  L_0 T_0 \Bigg[ e^{i \omega \frac{T_0}{2}} sinc(\omega \frac{T_0}{2}) \ast_\omega \delta( \omega - 2 \omega_{0} ) \ast_\omega 
     \\ &\qquad 
     e^{-\frac{\omega^2}{4\sigma^2}} \left( e^{ik\frac{L_0}{2}} sinc(k\frac{L_0}{2}) \ast_k \delta(k+2\gamma P) \ast_k \delta(k  + \omega^2 k'') \right)_{k=\Delta  k(\omega_s,\omega_i)}  
     \Bigg]_{\omega=\omega_s+\omega_i} 
    \label{eq:4wavemix}
    \end{split}
\end{align}

The domain of convolution is specified with a subscript, and~the distributions are evaluated at the given expressions of signal and idler~frequencies. 

To compare Equation~(\ref{eq:4wavemix}) to the standard result, Equation~(\ref{eqn:JSF-LI}), 
we assume 
%$k''$ to be negligible, and assume 
asymptotic time ($T_0 \ra \infty$), in~which case $sinc(\omega T_0) \ra \delta(\omega)$, so the  convolution over $\omega$ reduces to the identity operation, and~we assume that the pump photon dispersion, $\omega^2 k''$, is negligible. Under~these conditions,
\begin{align*}
    \bs
    F(\omega_s, \omega_i) &\approx L_0 \Big( \delta( \omega - 2 \omega_{0} ) \ast_\omega  e^{-\frac{\omega^2}{4\sigma^2}} \left( e^{i(\Delta k+2\gamma P)\frac{L_0}{2}} sinc((\Delta k+2\gamma P)\frac{L_0}{2}) \right) \Big)_{\omega=\omega_s+\omega_i}     \\
    &\approx L_0   e^{-\frac{(\omega_s+\omega_i - 2\omega_{0})^2}{4\sigma^2}} \left( e^{i(\Delta k+2\gamma P)\frac{L_0}{2}} sinc((\Delta k+2\gamma P)\frac{L_0}{2}) \right) 
    \es
\end{align*}
which matches Equation~(\ref{eqn:JSF-LI}).

The newly-derived expression allows for short-time calculations using easily computable routines. In~place of a radical in the denominator, the~higher order effects of the pump are incorporated into the convolution~operations.
 
\subsection{Quantum Zeno~Dynamics}
The quantum Zeno effect would be a potentially fruitful experimental case to verify the utility of the new method for calculating second order amplitudes. This effect can be used, for~instance, to~generate entanglement by suppressing first-order terms relative to second-order terms in the Schrodinger Equation \cite{Nodurft2022}. Nodurft et al. use Fermi's Golden Rule to demonstrate their scheme utilizing a photonic waveguide coupled with a series of perturbing atoms to eliminate outcomes in which both photons appear at the same output port. This effectively entangles the photons by making the wave function inseparable. They write, ``this imbalance between single and two photon absorption rates facilitates the Zeno effect by suppressing terms where both a left and right photon are found, but~not when a single photon is found.'' Performing this calculation to higher precision using the RFT method presented here may provide more control and efficiency in creating entangled states. It would be interesting to use as a test case for verification that this method (see Sections~\ref{hdr:FGR1} and \ref{hdr:FGR2}) does indeed improve upon Fermi's Golden~Rule.

Other relevant potential applications of this calculation applied to the Zeno effect include error correction in quantum computers~\cite{EREZ2004} and construction of quantum gates~\cite{FRANSON2004}.

\subsection{Solitons and Non-Linear~Cases}
The method provided here can be directly applied to any dynamical case in which the potential function has a Fourier transform.  Its generality may make it particularly helpful for nonlinear versions of the Schrodinger equation. As~an example, Bayindir et. al. study q-deformed Rosen--Morse potentials to analyze the time evolution of solitons~\cite{BAYINDIR2021}. They utilize a spectral method to determine the stationary states of the system and~the Runge--Kutta (RK) method to evolve the stationary states in~time.

Non-linear terms are hard to handle because they involve convolution in the frequency domain, due to the convolution theorem. One typically avoids this by iterating them in the spatial domain, whereas the linear terms are iterated in the frequency domain. In~the method presented here, the~convolutions in the frequency domain are taken into account explicitly, rendering the spatial iteration of the nonlinear terms unnecessary. It seems that this method could be used as an alternative to RK to compute time evolution, readily handling linear and nonlinear~terms. 

It should be noted that the RK method results in a time domain wave function, whereas the methods here result in a frequency domain wave function. These can be easily compared using forward or inverse Fourier~transforms.

\subsection{Master~Equations}
When considering open quantum systems, the~Lindblad equation can be used to account for dissipative or decoherence effects from the system to the environment. These effects are accomplished by a nonlinear term added to the Schrodinger equation, requiring careful treatment. There is still, however, a~time integration that must be accomplished, which is often done using the Runge--Kutta method. The~alternate method proposed here can likely be applied to that integration process, providing some utility in these cases. A~careful analysis of these benefits was not~undertaken.

\subsection{Other~Applications}
The method proposed based upon Fourier transforms has a quite general form and might be used in other scenarios. The~TDSE describes the diffraction of the wave function around a small temporal perturbation. One might consider application in the spatial domain, rederiving the usual single slit diffraction formula, and~then extending this to a second-order calculation. Furthermore, in the spatial domain, one might apply the RFT technique to a tunneling barrier, for~instance in a scanning tunneling microscope or in the alpha~decay.

\section{Summary} \label{summary}
In this work, we devised a novel technique that decouples nested integrals in the Dyson series for the time-dependent Schr\"{o}dinger Equation (TDSE) using recursive Fourier transforms (RFT). This provides an approach which is particularly suited for computation on both classical and quantum~computers.

This method shares similarities with existing multi-slice or split-operator techniques, but it~is used to refine accuracy of wavefunction spectra rather than propagate a wavefunction over time. The~RFT approach computes the temporal diffraction of a wavefunction under a perturbing force of finite duration. It can be used, for~instance, in~the characterization of single photons in cases where indistinguishability is~important.

The decoupling of the integrals at second-order is achieved by shifting to the frequency domain to obtain a nested $sinc$ function, then interpreting the nested $sinc$ as a function of time while also swapping the order of operators to perform the outer time integral before the sum over energy. This varies the width of the $sinc$ function in the frequency domain, which can be sampled at a given frequency to extract an amplitude in the frequency domain. This allows the TDSE to be expressed as a sum of (non-nested) convolutions. We anticipate that this procedure can be iterated to higher~orders.

%%%%%%%%%%%%%%%%%%%%%%%%%%%%%%%%%%%%%%%%%%
\vspace{6pt} 

%%%%%%%%%%%%%%%%%%%%%%%%%%%%%%%%%%%%%%%%%%
%% optional
%\supplementary{The following supporting information can be downloaded at:  \linksupplementary{s1}, Figure S1: title; Table S1: title; Video S1: title.}


\funding{This research received no external~funding.}

\dataavailability{The author declares that there are no data associated with this paper. Simulation code to generate figures can be supplied upon request.}

\acknowledgments{The author is grateful to Jeff Butler, Richard Pham Vo, Marcin Nowakowski, Eliahu Cohen, Paul Borrill, Andrei Vazhnov, Stefano Gottardi, Daniel Sheehan, Joe Schindler, Mark Prusten, and Justin Kader for helpful comments and~feedback.
}

\conflictsofinterest{The author declares no conflicts of~interest.} 

%%%%%%%%%%%%%%%%%%%%%%%%%%%%%%%%%%%%%%%%%%
\abbreviations{Abbreviations}{
The following abbreviations are used in this manuscript:\\

\noindent 
\begin{tabular}{@{}ll}
MDPI & Multidisciplinary Digital Publishing Institute\\
RFT & Recursive Fourier transform\\
TDSE & time-dependent Schr\"{o}dinger equation\\
\end{tabular}
}

%%%%%%%%%%%%%%%%%%%%%%%%%%%%%%%%%%%%%%%%%%
%% Optional
\appendixtitles{yes} % Leave argument "no" if all appendix headings stay EMPTY (then no dot is printed after "Appendix A"). If~the appendix sections contain a heading then change the argument to "yes".
\appendixstart
\appendix

\section[\appendixname~\thesection]{Definitions of the Fourier Transform} \label{hdr:FT_def}
The following definitions of the Fourier transform and its inverse are used.
\begin{align*}
    F(\omega) = \frac{1}{\sqrt{2\pi}} \int_{-\infty}^{\infty} f(t) e^{i\omega t} \, dt 
    \qquad \qquad 
f(t) = \frac{1}{\sqrt{2\pi}} \int_{-\infty}^{\infty} F(\omega) e^{-i\omega t} \, d\omega
\end{align*}
\begin{align*}
    F(k) =  \frac{1}{\sqrt{2\pi}} \int_{-\infty}^{\infty} f(x) e^{-i k x} \, dx
\qquad \qquad
f(x) =  \frac{1}{\sqrt{2\pi}}  \int_{-\infty}^{\infty} F(k) e^{i k x} \, dk
\end{align*}

\section[\appendixname~\thesection]{Using the Appropriate Dual Domain} \label{hdr:domainExamples}

%\section{Appendix \label{hdr:appendix}}

%\subsection{Using the appropriate dual domain \label{hdr:domainExamples}}
%Many problems can be expressed equivalently in either of two dual domains, such as time and energy, or space and momenta. Because some problems are easier to solve or are more intuitive to understand in one domain than the other, during calculations one should consider what the natural domain is.
%For instance, in the momentum domain the momentum $p$ is explicit whereas in the position domain, momentum is implicit ($\frac{d}{dx}$).

In systems linked by Fourier transforms, a~proper domain emphasis can sometimes be overlooked. For~example, TDSE coefficients $c(t)$ are written as functions of time to establish the time dependence of the~wavefunction.

However, Equation \eqref{eqn:mainResult1} (Figure \ref{fig:Gaussian_Potential}b) represents a $\om$-space distribution featuring an $\om$ convolution. Time does not appear directly in this expression. Instead, $\Time$ is a constant that shapes the oscillatory pattern of the distribution at a given moment. By~varying $\Time$, we must recompute the convolution at each time step and then sample the distribution at point $\omega_f$ to yield a meaningful transition amplitude. This requires distinguishing ``integration parameters'' from ``coordinates'', in~the sense of~\cite{NELSONISAACS2021}.

\begin{figure}[H]
 
		\hspace{-4pt}\includegraphics[width=\linewidth]{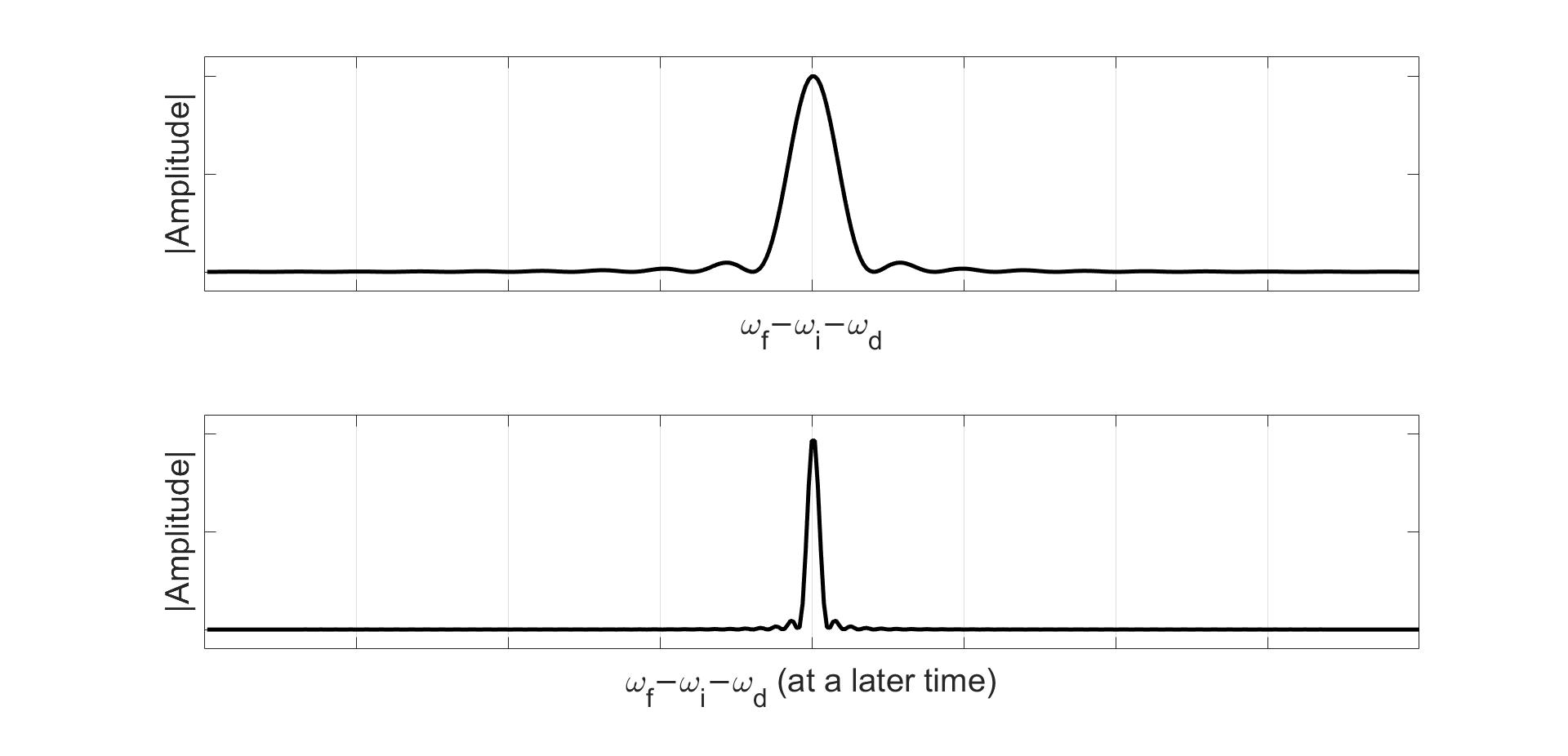}
		\caption{Fermi's Golden Rule, transition amplitude. There is more than one possible interpretation for this plot. (\textbf{Top}) For a given driving frequency, we treat the energy jump $\omega_{fi}$ as the variable. The~greatest amplitude occurs when $\omega_{fi}$ matches the driving frequency, $\omega_{d}$. (\textbf{Bottom}) At a later time, the resonance curve has changed, so this can serve to predict time evolution despite that it is a distribution in $\om$ space.
		} \label{fig:fermiGraph1}
\end{figure}

Consider Fermi's Golden Rule: a bound state $\omega_i$ transitions to a continuum state $\omega_f$ under a driving frequency $\omega_d$, producing the transition amplitude expressed in Equation~\eqref{eqn:FermiPerturbCoeffOld}. By~varying $\omega_d$, $\omega_i$, or~$\omega_f$, Figure~\ref{fig:fermiGraph1} is useful for identifying the relevant~dependencies.

However, Figure \ref{fig:fermiGraph1} can also be interpreted time-wise because the $sinc$ function depends  symmetrically on time and energy. Over~time, the~$sinc$ function (as a function of $\omega_{fi}-\omega_d$) becomes more peaked, and~the image in Figure \ref{fig:fermiGraph1} is considered to be a snapshot of time. Thus, we interpret the amplitude as time-dependent, $c(t)$.

However, this interpretation misreads the proper domain. Amplitude $c_{fi}$ is a frequency distribution and not a time distribution. The~time dependence is implicit; evolving time means updating the entire distribution, after~which we can derive the frequency-dependent amplitude at that~time. 

These processes are distinct. The~former involves only number reading from the graph, while the latter requires repeated graphing and sampling. The~former is a function, whereas the latter is a~functional.

The same reasoning applies to the usual kicked harmonic oscillator treatment (perturbed by a small Gaussian pulse, Section $\ref{hdr:TDSE1stOrder}$). The~standard methods lead to the coefficient in Equation \eqref{eqn:gaussKickB}, which is implicitly defined by the elapsed time $\Time$ but explicitly a function of $\om$. This can help determine the best pulse duration $\Time$ to match the natural oscillator frequency $\om$, but~it overshadows the more natural $c = c(\om)$ dependence.

Each $\DeltaT$ value is a unique experiment leading to a different distribution. In~varying $\DeltaT$, Equation \eqref{eqn:gaussKickB} becomes a functional by creating a configuration space for each $\DeltaT$ value.

%\subsection[\appendixname~\thesubsection]{}

%\section[\appendixname~\thesection]{}
\section{Detailed~Analysis} \label{hdr:funcanalysis}
%\subsection{The second-order transfer function $\PSIB$}
We will now examine Equations \eqref{eqn:mainResult2} and \eqref{eqn:mainResult3} in more~detail.

\subsection{Evaluating the Transfer~Function} \label{hdr:PSIB}
Evaluating the form of the transfer function $\PSIB$ can be performed first for the case of a negligible potential, $\potV = \delta(\om)$ (infinitely wide in the time domain).
Because convolution with a $\delta$-function is the identity operation, we can then evaluate $\psib[t]$ explicitly and take its Fourier transform,
\begin{align}
    \label{eqn:H_w1}
    \begin{split}
    \PSIB &= \underset{\DeltaT_1 \ra \om}{\scrT}\Big\{ \exp{(i\omega_{ki} \DeltaT_1/2)} \frac{sin(\omega_{ki} \DeltaT_1/2)}{\omega_{ki}/2} \Big\} \\
    &= \frac{2\pi i}{\omega_{ki}} \left(\delta(\om + \omega_{ki}/2 + \omega_{ki}/2) - \delta(\om - \omega_{ki}/2 + \omega_{ki}/2) \right) \\
    &= \frac{2\pi i}{\omega_{ki}} \left( \delta(\om + \omega_{ki}) - \delta(\om) \right).
\end{split}
\end{align}

In other words, the~transfer function is composed of a series of discrete impulses spaced at integer multiples of $\groundom$ (because $\omega_{ki} = k \groundom$). See Figure~\ref{fig:mainresult2}.

Next, $\PSIB$ is convolved with the other factors in Equation \eqref{eqn:mainResult3}, resulting in
\begin{align}
    \label{eqn:shifted_impulses}
    \begin{split}
    \delta(\om - \omega_k) \ast \PSIB &= \frac{2\pi i}{\omega_{ki}}\delta(\om - \omega_k)  \ast \left( \delta(\om + \omega_{ki}) - \delta(\om) \right) \\
    &= \frac{2\pi i}{\omega_{ki}}\left(\delta(\om - \omega_i)  - \delta(\om - \omega_k)\right)
\end{split}
\end{align}

As shown in Figures~\ref{fig:Hw_impulses_1} and \ref{fig:painting1}, each term in the sum over $k$ contributes complex impulses at $\omega_k$ and $\omega_i$. 

The special case $k=i$ must be handled separately. Here, the~desirable properties of the $sinc$ function at the origin are required, and~Equation \eqref{eqn:H_w1} is the Fourier transform of a~constant as follows:
\begin{align}
    \label{eqn:H_w2}
    \begin{split}
    \tilde{H}_{00} &= \scrT\Big\{ \text{constant} \Big\} = 2\pi \delta(\om).
\end{split}
\end{align}

This contributed to a purely real amplitude at the origin, as~shown in the middle graph of Figure~\ref{fig:Hw_impulses_1}.

Summing over all $\PSIB$ contributions over the range $k=\pm25$, it can be seen in Figure~\ref{fig:Hw_impulses_2} that for $k\neq 0$, $\PSIB$ contributes a real portion at $\omega_k=0$ which is amplified as more frequencies are included ($k\rightarrow \pm \infty$), whereas the real portion remains small and finite at every other $\omega_k$, vanishing when the distribution is normalized (top). Conversely, the~imaginary portions cancel at $\omega_k=0$ but are significant everywhere else, decaying inversely with respect to $|k|$ (middle). 
%These impulses are convolved with the first-order impulse response, Equation $\ref{eqn:mainResult1}$, and evaluated at $\omega_f$ to obtain the coefficient $\cnats$.

An analogy can be drawn to the frequency-domain decomposition of sound signals. In~the second-order calculation, the~probability amplitude signal was deflected into a series of higher harmonics. Similarly, an~acoustic musical instrument generates sound through the combination of a pluck (impulse) and resonant cavity that amplifies higher harmonics (impulse response). This is similar to the relationship between $\PSIB$ in Equation \eqref{eqn:mainResult2} and the rest of that~equation.

\subsection{Stepping Through the Algorithm for Transfer~Function}

\vspace{-3pt}
\begin{figure}[H]	
	 
		\hspace{-3pt}\includegraphics[width=\linewidth]{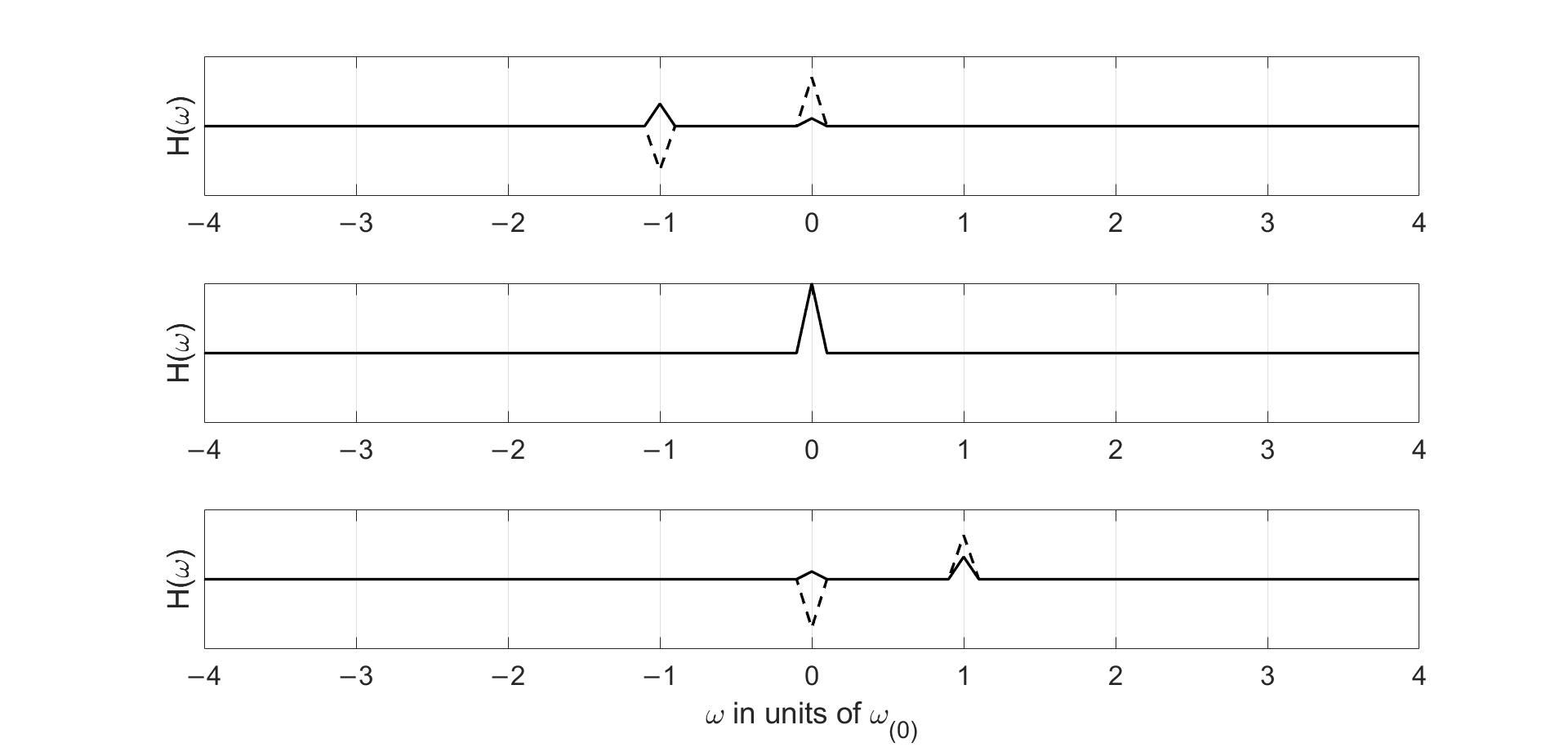}
		\caption{\textls[-15]{Contributions to the second-order complex valued ``transfer function'' $\PSIB$ from the frequencies $k=-1,0,1$ (top to bottom, respectively). The~cases $k=\pm1$ contribute a real portion (solid) and an imaginary portion (dashed) at $\omega_k=\pm1$. The~case $k=0$ contributes only a real portion (solid). See Equations \eqref{eqn:H_w1} and \eqref{eqn:H_w2}}.} \label{fig:Hw_impulses_1}
\end{figure}
\unskip

\begin{figure}[H]	
	 
	   \hspace{-3pt} \includegraphics[width=0.95\linewidth]{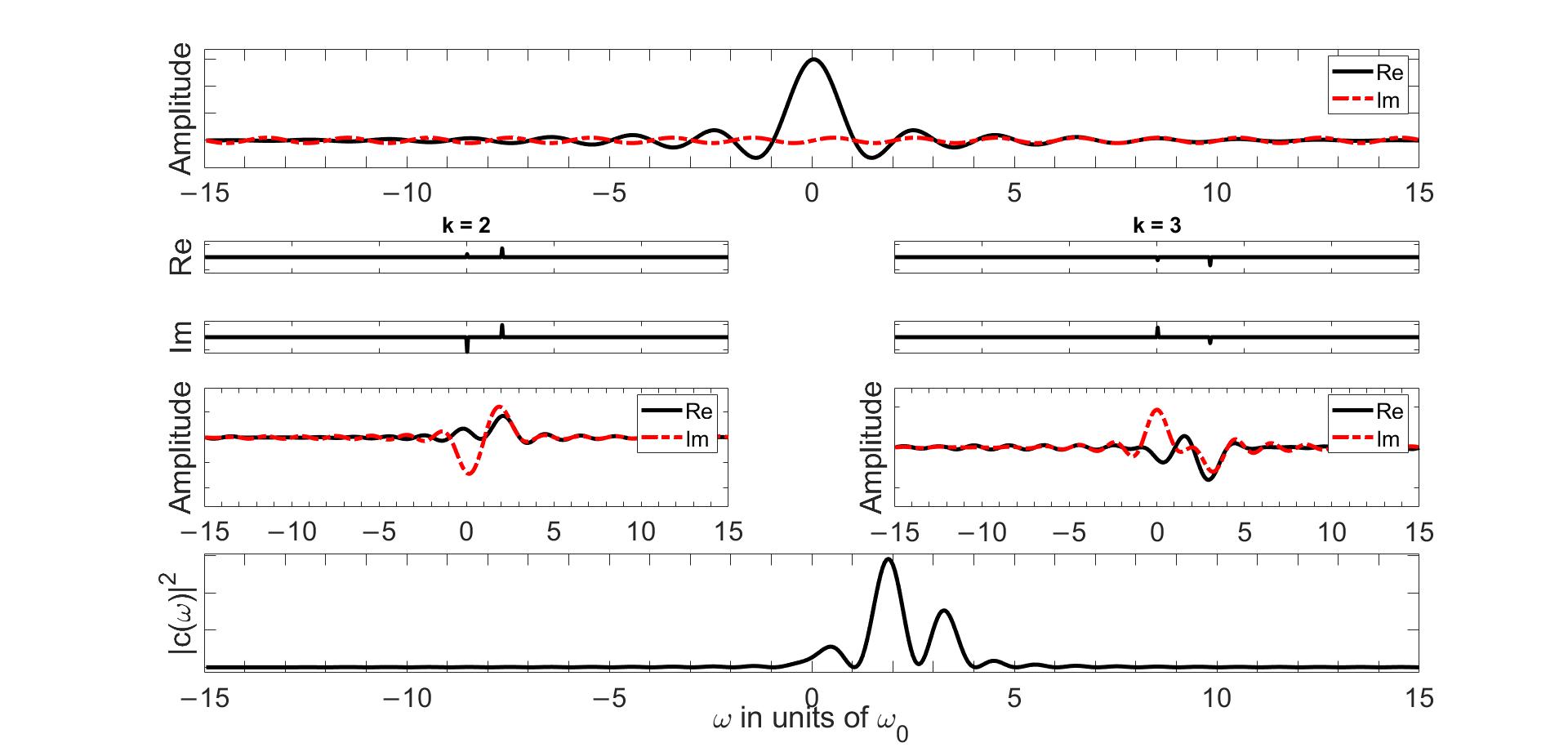}
		\caption{The frequency-distributions for $k=2,3$ contributions. (\textbf{Top}) The real (solid) and imaginary (dashed) parts of the complex $sinc$ function in Equation \eqref{eqn:mainResult2}, corresponding to the outer integral over $t_1$ in Equation \eqref{eqn:2ndOrder_step5}. (2nd/3rd row) The transfer function $\PSIB (\om)$ in Equation \eqref{eqn:mainResult3} captures information from the nested $t_2$ integral as a series of spikes, $\delta(\om-2\groundom)+\delta(\om)$, (\textbf{left}), $\delta(\om-3\groundom)+\delta(\om)$, (\textbf{right}). See Equation \eqref{eqn:H_w1}. (4th row) The real (solid) and imaginary (dashed) parts of the convolution in Equation \eqref{eqn:mainResult2}. (\textbf{Bottom}) The combined result from $k=2$ and $k=3$.} \label{fig:painting1}
\end{figure}

A comparison was made between the direct integration of Equation \eqref{eqn:DiracTDSE2} and the convolution approach in Equations~(\ref{eqn:mainResult2}) and \eqref{eqn:mainResult3} using MATLAB.
The program begins by generating a nested impulse response, Equation \eqref{eqn:impulseResponse}. This has the form $sinc \ast V$ and width $\DeltaT_1$, as~shown in Figure~\ref{fig:mainresult2} (far-left). 

This impulse response (in $\om$) is then \textit{varied over time} across the integration limits $0<\DeltaT_1<\Time$, generating a sequence of $sinc$ graphs of varying widths. The~sample value at the vertical line $\omega_k$ for each graph was stored as a new array $\psib[\DeltaT]$ (Figure \ref{fig:mainresult2}, middle). For~a given $\omega_k$, these samples oscillate with a frequency profile that is dependent on the physics of the experiment (such as the properties of the potential and duration of the window of measurement).

\begin{figure}[H]	
	 
		\hspace{-2pt}\includegraphics[width=\linewidth]{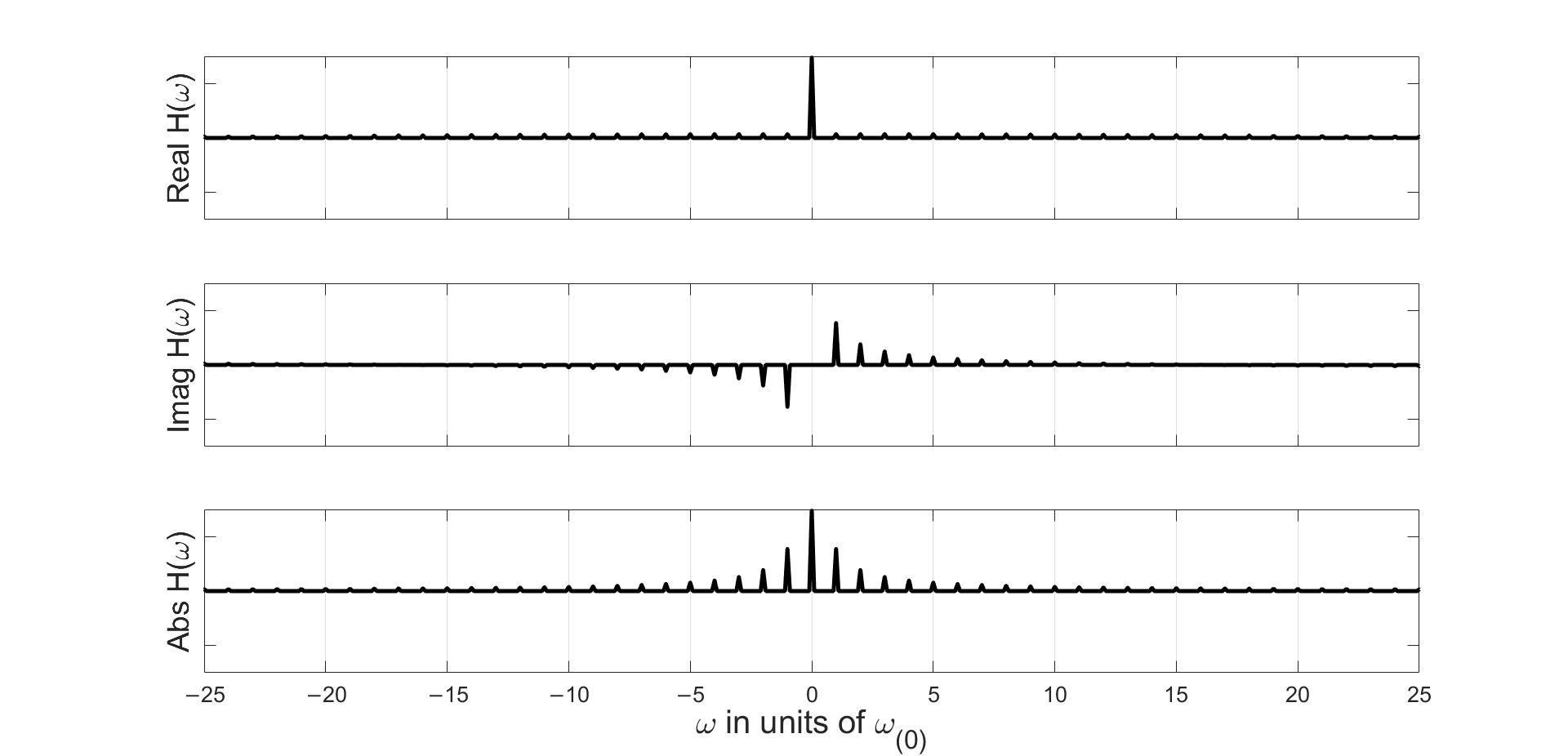}
		\caption{$\PSIB(\om)$ for $-25<k<25$. The~contribution to $\PSIB(\om)$ at $k=0$ is non-zero for every $k\neq0$, growing without bound, as~we include more momenta ($k\rightarrow \infty$), see Equation \eqref{eqn:H_w2}. There is also a small, constant, real contribution at each $\omega_k$, which vanishes when the distribution is normalized (\textbf{top}). The~imaginary portions cancel at $\omega_k=0$ and are built everywhere else, decaying inversely to $|k|$ (\textbf{middle}), see Equation \eqref{eqn:H_w1}. The~absolute magnitude of $\PSIB$ was a harmonic series of impulses (\textbf{bottom}). These impulses are convolved with the first-order impulse response, Equation \eqref{eqn:mainResult1}.} \label{fig:Hw_impulses_2}
\end{figure}

The Fourier transform of $\psib[\DeltaT]$ is  $\PSIB(\om)$ (Figure \ref{fig:mainresult2} right panel). It is a series of impulses representing each intermediate contribution to the second-order amplitude (see \mbox{Figures~\ref{fig:Hw_impulses_1} and \ref{fig:Hw_impulses_2}}). If~the perturbation is negligible, we can write $\tilde{V}(\om)=\delta(\om)$. In~this case, $\PSIB$ is composed of $\delta$-function impulses at $\omega_k$ and $\omega_i$. If~the potential is strong, other harmonics appear in this graph (see Figure~\ref{fig:H_padding1} bottom-right).

Finally, in~Equation \eqref{eqn:mainResult2}, $\PSIB$ is convolved with a phase-shifted $sinc$ impulse response so that a copy of the impulse response is placed wherever $\PSIB$ has a spike, as~shown in Figure~\ref{fig:painting1}.
This is performed for every possible intermediate state $\omega_k$, and~the amplitude plots for each are summed. Each code loop over $\omega_k$ contributes an impulse response centered at $\omega_k$ and another centered at $\omega_i$ (Figure \ref{fig:Hw_impulses_1}). 

After looping over all $2k+1$ intermediate distributions, the~impulses centered on $\omega_i$ reinforce $2k+1$ times, whereas the second-order signal at each $\omega_k$ appears only once. The~result is a strong central peak and~decaying wings (Figure \ref{fig:ordercompare1}).

\subsection{Domain and Resolution of Transfer~Function}
\begin{figure}[H]	
	 
		\hspace{-3pt}\includegraphics[width=\linewidth]{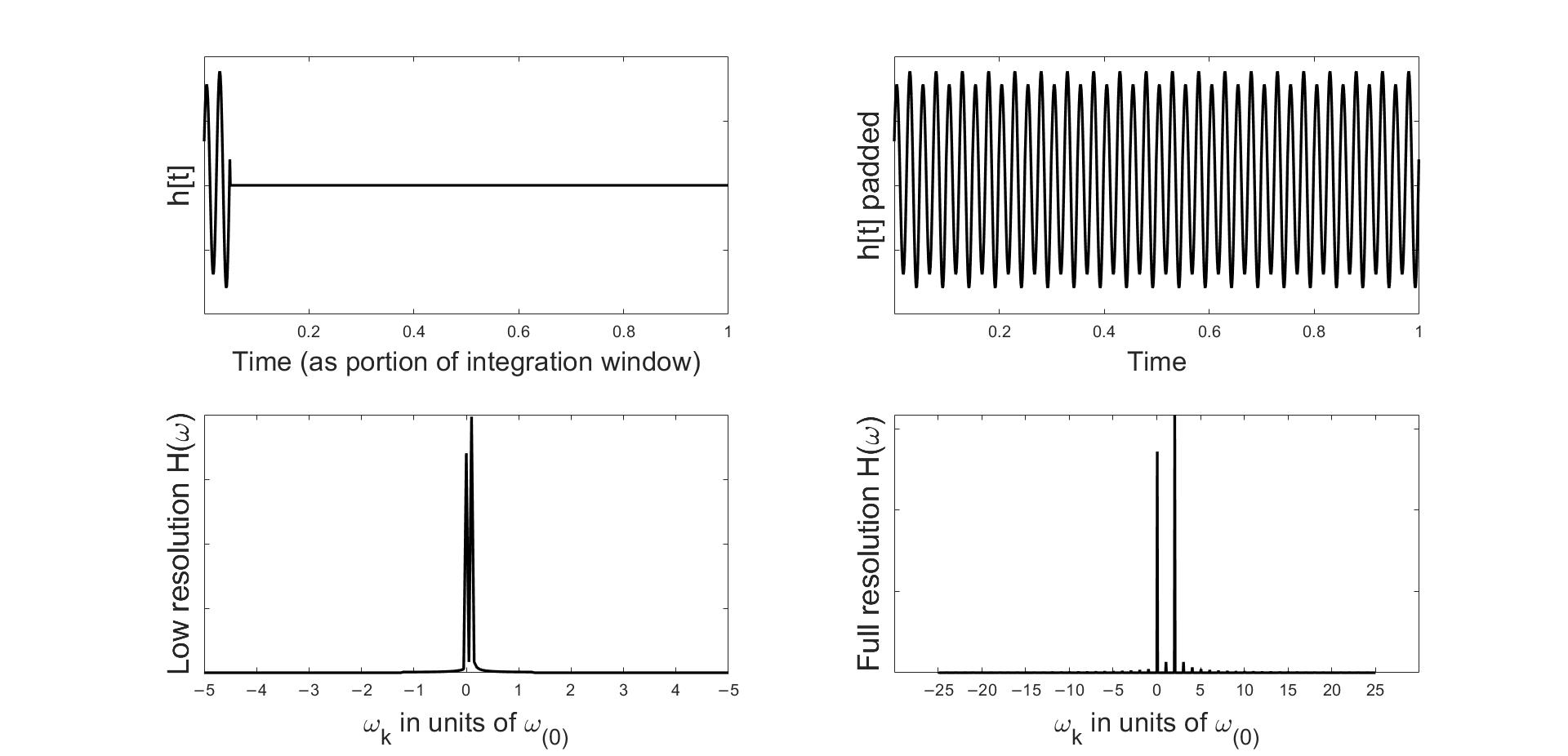}
		\caption{Bandlimited signal: Because the duration of the time integration of the TDSE is less than the full signal (5\% shown here, \textbf{top left}), $\psib[t]$ is smaller than $\tilde{V}$ by that factor, and~the resolution is decreased (\textbf{lower left}, the~distinct spikes are only resolvable because horizontal scale is expanded). In~the \textbf{upper right}, $\psib[t]$ was padded with copies of itself (\textbf{upper right}) to compensate for the limited bandwidth of the signal. This ensures its Fourier transform has the desired high resolution (\textbf{lower right}). Shown here is the $k=+2$ term of Equation $\ref{eqn:mainResult3}$. A~moderate strength potential was used, resulting in harmonics at nearby states (see small spikes at $k=+1, +3$ and other integers in \textbf{bottom~right}). } \label{fig:H_padding1}
\end{figure}

%It is to be expected for any bandlimited signal that its spectrum will be periodic.
%THIS IS SIMPLY THE USUAL BANDLIMITED SIGNAL  THAT IS REQUIRED FOR THE SAMPLING THEOREM. WE SAMPLE THE SIGNAL OVER A LIMITED RANGE, AND ASSUME IT IS PERIODIC WITH ALL FREQUENCIES EXISTING WITHIN A CERTAIN RANGE.
In the code implementation of Equation \eqref{eqn:impulseResponse}, the~length of $\psib[\DeltaT_1]$ is not equal to the length of the original signal. This is because $\psib[\DeltaT_1]$ is generated by scanning $\DeltaT_2$ over the variable range $\tLowerLimit<\DeltaT_2<\tUpperLimit$.
Thus, the~corresponding resolution of its Fourier transform, $\PSIB$, is reduced
%by a factor of the relative lengths of the arrays. 
(Figure \ref{fig:H_padding1}, top left). 

To compensate for the band-limited spectrum, $\PSIB$ was padded with copies of itself. This is necessary for the convolution operation to be well-defined. The~time window $\TUpperLimit$ was chosen to be an integer fraction of the duration of the original signal so that the padding fits evenly (this is necessary to avoid artifacts in the Fourier transform). 

This defines a fundamental harmonic frequency associated with the measurement,
\begin{align}
    \begin{split}
    \fundom &\equiv \text{total duration of original signal}/\text{duration of time integration}\\
    &= \frac{1}{\Time},
    \end{split}
\end{align}
(see the harmonic spacing in Figure~\ref{fig:Hw_impulses_2}).

When the Gaussian potential is weak, the~tiled instances of $\psib[\DeltaT_1]$ line up smoothly, and~$\PSIB$ only contains two spikes, as~in Equation \eqref{eqn:H_w1} and Figure~\ref{fig:Hw_impulses_1}.
When the potential is stronger, $\psib[\DeltaT_1]$ does not line up on its endpoints, and~spectral artifacts occur at integer multiples of $\fundom$.

For reasons that are not fully understood by the author, the~interpretation of the second-order results is clear only when $\fundom = \groundom$, which is known as cyclotron resonance.
This appears to be related to the interpretation of Equations~(\ref{eqn:mainResult2}) and \eqref{eqn:mainResult3} as a signal reconstruction problem using sinc-interpolation: this is the only case considered in this~study.

\subsection{Effect of Time Window Shift on the Form of the Transfer~Function}

In Figure~\ref{fig:Hw_impulses_2}, the~imaginary part of $\PSIB$ decays inversely with $|k|$. The~time measurement window was shown to extend from the origin to $\DeltaT$, leading to a translational factor of $\frac{1}{2}$ in the $rect(t)$ function. In~the general case, the~measurement window can be translated $r$ units by shifting the $rect(t)$ function again, $rect(\frac{t}{\DeltaT}-\frac{1}{2}-\frac{r}{\DeltaT})$, leading to an overall phase shift in the frequency domain, $\exp{(i \omega_k (r/\DeltaT))}$. This leads to an oscillating envelope for the impulses in Figure \ref{fig:Hw_impulses_2} (envelope not shown). In~the MATLAB simulation, a phase shift of this sort was introduced to compensate for coding artifacts (the base index for the time window started at 1 instead of 0).

\subsection{Normalizability of the Transfer~Function}\label{hdr:norm}
The appearance of $\omega_{ki}$ in the denominator of Equation $\eqref{eqn:impulseResponse}$ inside summation over both $i$ and $k$ is the cause for questioning whether this expression can be normalized. However, owing to the good properties of the $sinc$ function, $\psib$ and $\PSIB$ are non-singular.
%This is a path integral through the intermediate energy eigenstates.

To observe this, note that when $1/\omega_{ki}$ becomes singular, we use Equation \eqref{eqn:H_w2} (which is well defined) instead of Equation \eqref{eqn:H_w1}.

In general, $\PSIB$ is a series of harmonics of spacing $\groundom$, as~shown in Figure~\ref{fig:Hw_impulses_2}. The~middle plot shows imaginary impulses at every non-zero integer $\omega_k$ that form a harmonic series, which is well known to not converge, and therefore, it is not clear whether the final expression Equation  \eqref{eqn:mainResult2} is convergent. The~upper plot shows an impulse at $w_i=0$ resulting from each term in the sum over $k$. The~height of this impulse increases without bounds for $k\rightarrow \pm \infty$. This is ultraviolet~divergence. 

This can easily be resolved from a practical perspective. Because~the height of the impulse at the origin is proportional to the size of the domain, $|k_{max}|$, in~the code implementation, this expression can be normalized by dividing by the maximum value of $k$, where only a finite number of terms are~included. 

From a theoretical perspective regarding the convergence of the second-order, the~issue is whether the $sinc$ functions, each of which are normalizable and arranged in a harmonic series (which does not converge) are normalizable. This was not addressed in this~study.

\section{Accuracy of~Method} \label{hdr:erroranalysis}

To analyze the accuracy of Equations~(\ref{eqn:mainResult1}), \eqref{eqn:mainResult2} and \eqref{eqn:mainResult3}, we used MATLAB to compare the convolution method and the method of direct integration of the Schr\"{o}dinger equation in Equation \eqref{eqn:direct-integration}.

\subsection{Comparing First-Order to Second-Order~Convolution}

The first-order contributions and second-order corrections for the convolution (or recursive Fourier transform) method were compared, as~shown in Figure~\ref{fig:ordercompare1}. The~second-order contribution shortens the central peak, while heightening the wings of the distribution by a small amount. This is reasonable because we expect the potential to deflect the system away from its original state each time it is~applied. 

%Because the simulation runs over finite values of $\omega_k$, the edge of the final distribution near $\pm15$ shows an artifact due to its missing neighbors at the boundary. If the simulation was run across an infinite domain, this artifact would be removed.

\subsection{Frequency Profile versus Potential~Strength}

\vspace{-6pt}

\begin{figure}[H]
   
 \hspace{-10pt}    \begin{subfigure}[b]{0.33\textwidth}
         \centering
     		\includegraphics[width=\textwidth]{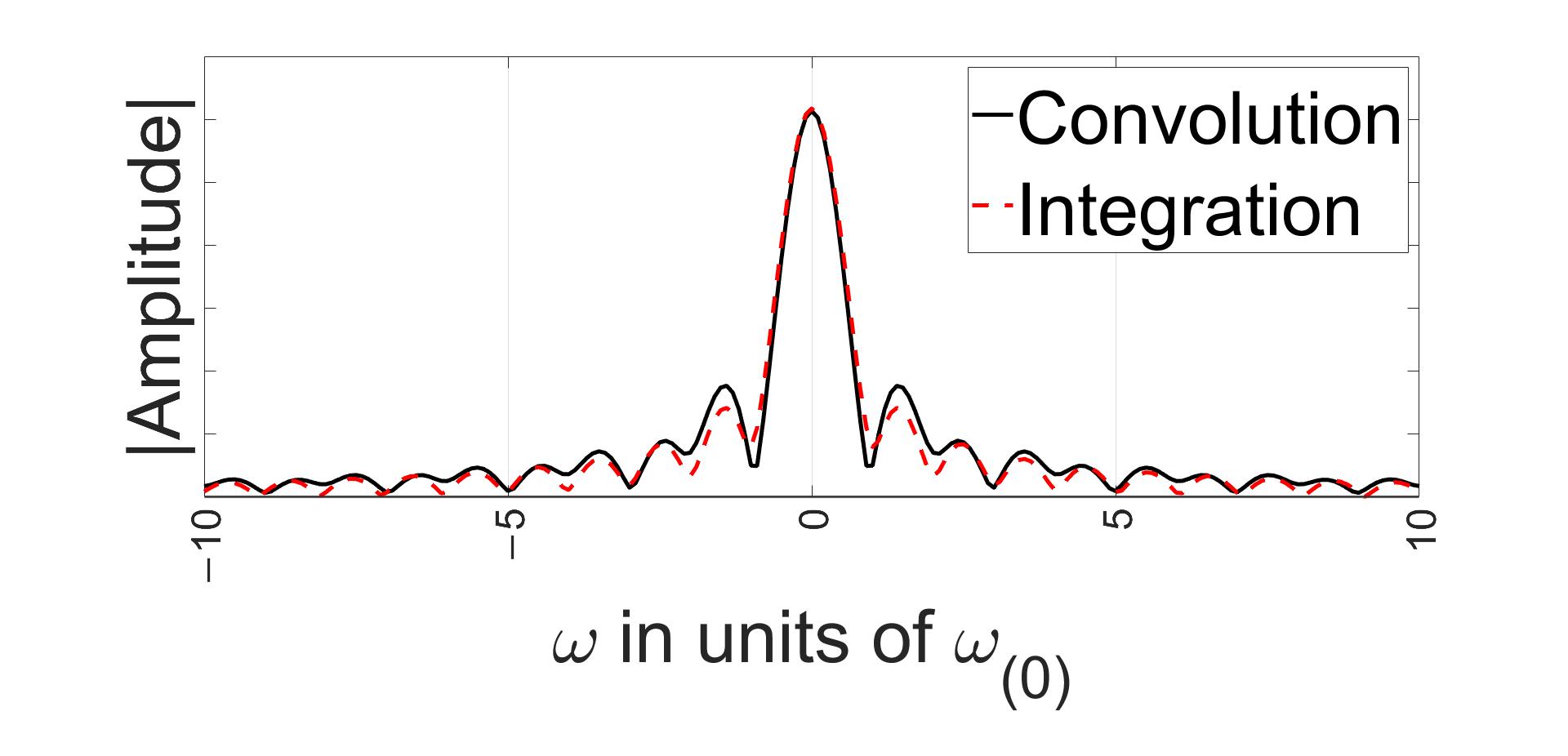}
     	%	\caption{Low~potential.}
         \label{fig:methodcompare-spectra-1}
     \end{subfigure}
    \begin{subfigure}[b]{0.33\textwidth}
         \centering
     		\includegraphics[width=\textwidth]{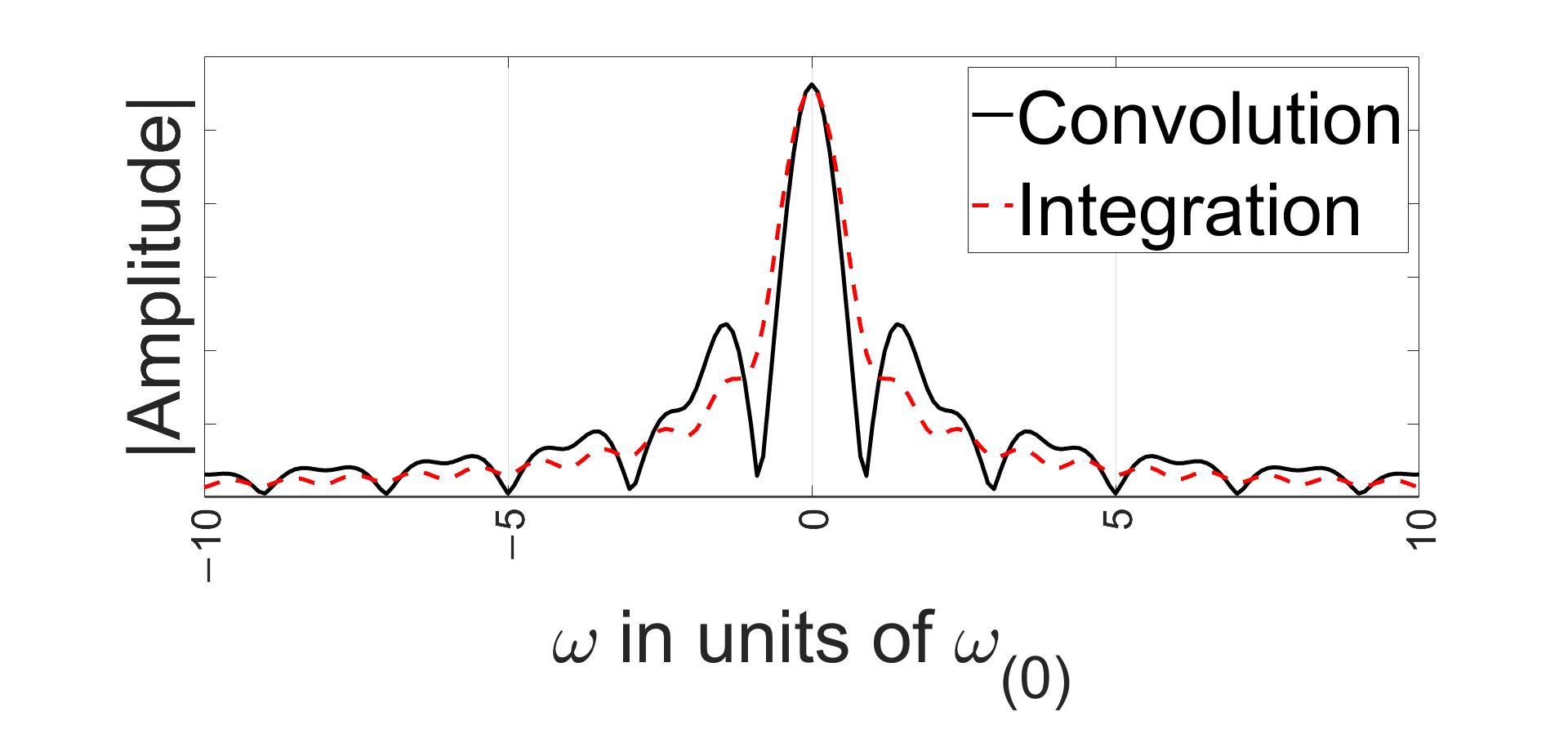}
     	%	\caption{Medium~potential.}
         \label{fig:methodcompare-spectra-2}
    \end{subfigure}
    \begin{subfigure}[b]{0.33\textwidth}
         \centering
     		\includegraphics[width=\textwidth]{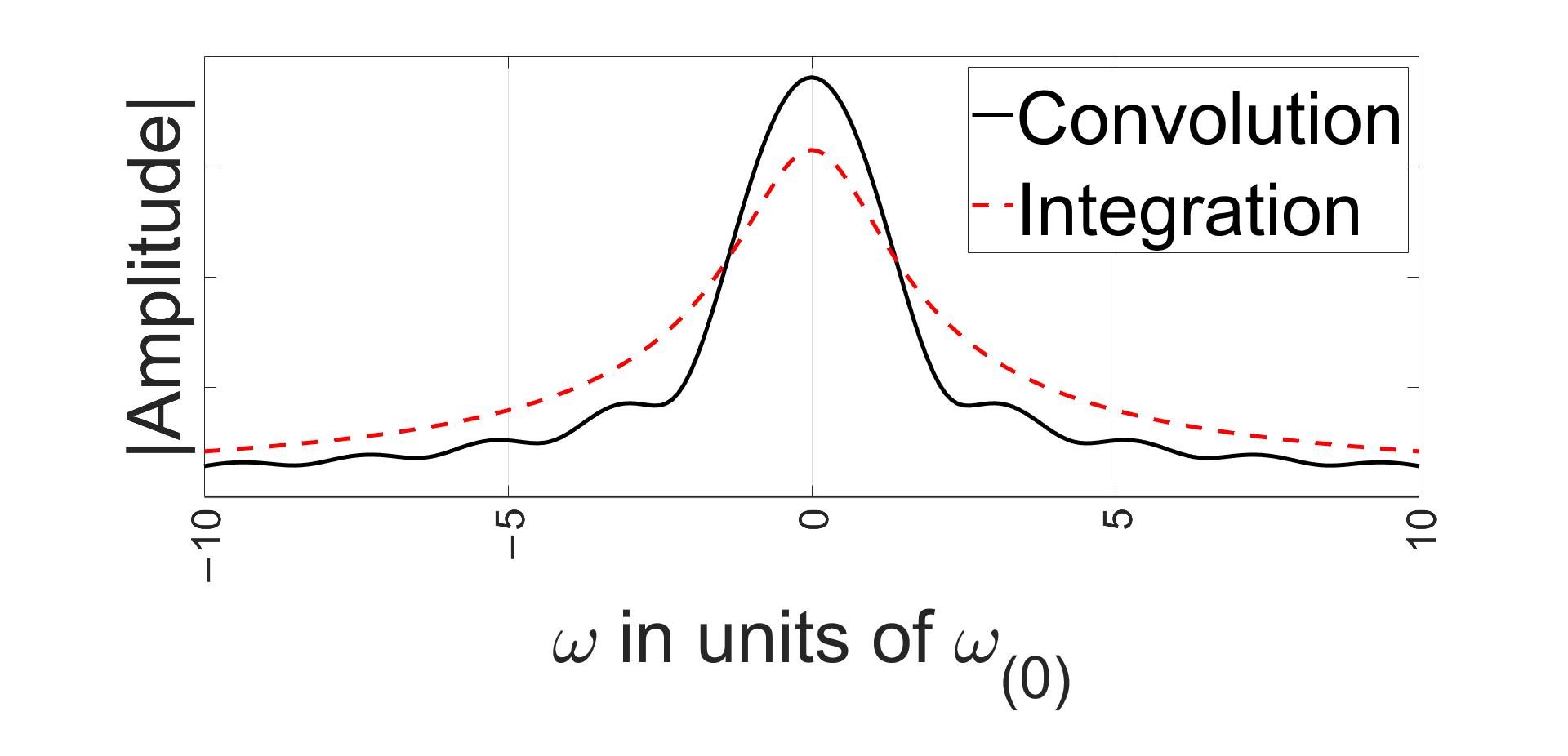}
     	%	\caption{High~potential.}
         \label{fig:methodcompare-spectra-3}
     \end{subfigure}\vspace{6pt}
     \caption{Second-order comparison of spectra for various values of the potential for convolution versus integration~methods. Potential increasing from left figure to right figure.}\label{a6}
\end{figure}

For the measurement of duration $\Time$, Figure~\ref{a6}a--c show the second-order amplitude calculated by direct integration and convolution. 
In both methods, increasing the potential spreads the central peak and smoothens the ripples in the distribution. For~a given increase in the potential strength, the~ripples are preserved to a greater extent in the convolution method. Furthermore, the~smoothing occurred differently in both cases. In~particular, Figure~\ref{a6}b shows that for the convolution method, the~odd-numbered zero-crossings are preserved longer than the even ones, as~the potential strength is~increased.

\subsection{Frequency Profile versus Range of Intermediate~States}

\vspace{-6pt}
\begin{figure}[H]
   
    \hspace{-12pt}  \begin{subfigure}[b]{0.45\textwidth}
            		\includegraphics[width=\textwidth]{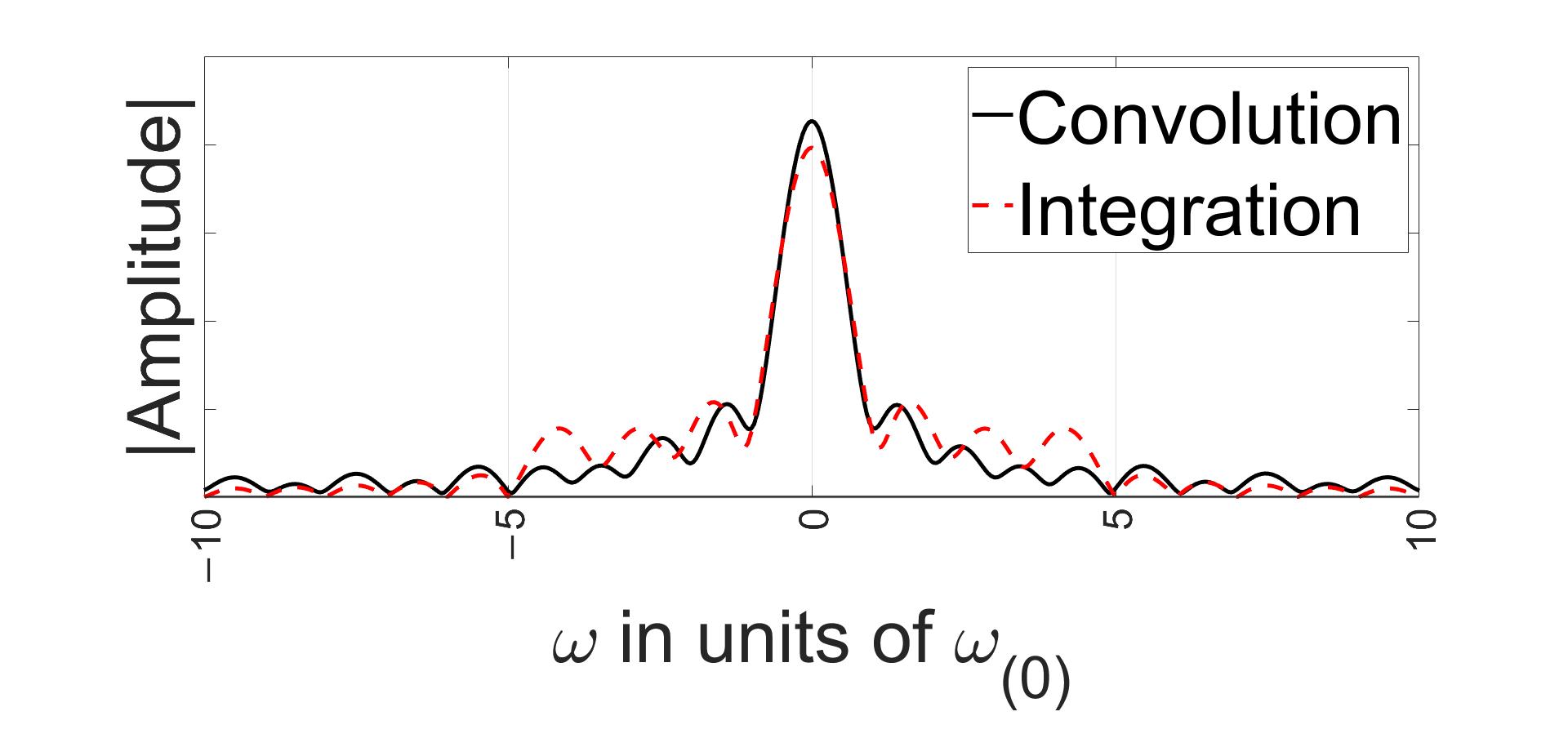}
      %  \caption{$k_{max}=\pm4$}
         \label{fig:methodcompare-range-1}
    \vspace{1em}
    \end{subfigure}
    \begin{subfigure}[b]{0.45\textwidth}
         \centering
		\includegraphics[width=\textwidth]{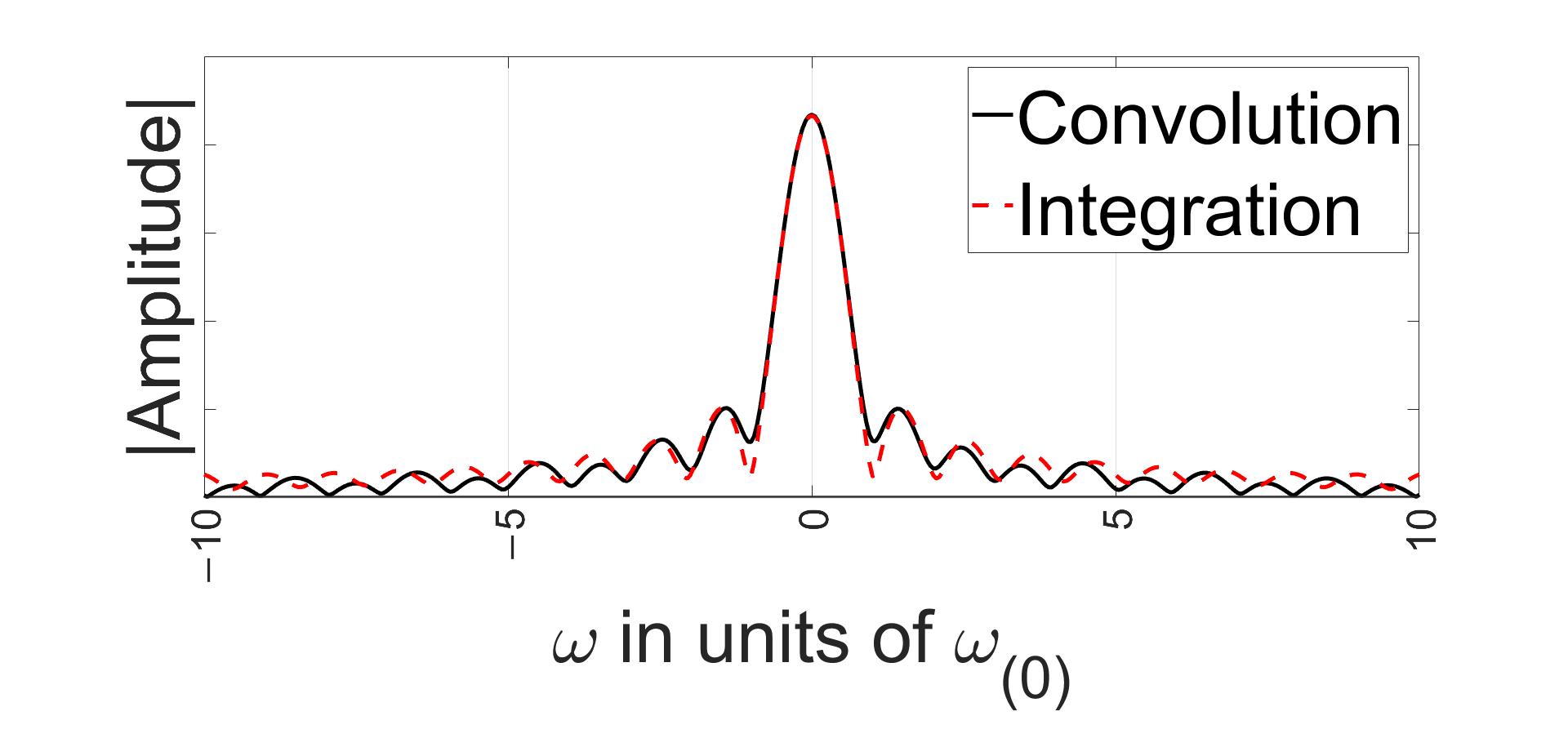}
     %	\caption{$k_{max}=\pm10$}
        \label{fig:methodcompare-range-2}
    \vspace{1em}
    \end{subfigure}
    \caption{Second-order comparison of spectra for two values of the $k_{max}$ range for convolution versus integration methods. The~methods match each other more closely far from the origin as more $k$ terms are~added. Left (right) figure case is $k_{max}=\pm4 (\pm10)$.}\label{aaa666}
\end{figure}

Figure~\ref{aaa666}a,b demonstrate the distinct behavior of direct integration versus convolution with respect to the number of intermediate states $k_{max}$ that are summed over. The~convolution method converges faster than the direct integration with respect to the number of intermediate terms included.
The convolution method relies heavily on non-local surrounding states. This is not surprising when considering the similarity between \mbox{Equations \eqref{eqn:mainResult2} and \eqref{eqn:mainResult3}} to the $sinc$ interpolation signal reconstruction (as shown in \mbox{Appendix~\ref{hdr:sampling}}).

\section{Interpretation}  \label{hdr:interpretation}
\unskip
\subsection{Kicked Frequency and Natural Frequency for Harmonic~Oscillator}

%Interpreting Eq. $\ref{eqn:mainResult1}$ requires care because the energy $\om$ appears as both a continuous and discrete element, as discussed in sec. $\ref{hdr:param-coord}$. To be rigorous in our interpretation, it is helpful to consider the example of a harmonic oscillator pe   rturbed by periodic kicks.
Equation \eqref{eqn:mainResult3} presents two distinct energy scales: the unperturbed harmonic oscillator's discrete spacing $\groundom$, and~the truncated perturbation-induced minima spacing $\fundom = 2\pi/\Time$. The~latter corresponds to the $sinc$ function zeros in Equation \eqref{eqn:mainResult1} and \eqref{eqn:mainResult3}. This truncated perturbation is comparable to the periodic kick potential in the Floquet theory, where the Hamiltonian splits into a free part (with a discrete eigenspectrum) and a kick part (with a continuous eigenspectrum).

The unitary evolution operator $\hat{U}$ comprises $\hat{U}_{free}$, which determines $\groundom$ and the unperturbed basis states, and~$\hat{U}_{kick}$, which establishes $\fundom$.

The dynamics of a Floquet system strongly hinge on the relationship between the natural frequencies of the two Hamiltonian parts: $\fundom$ and $\groundom$. Engels~\cite{ENGEL2007} investigated structured stochastic webs arising in the phase space for various energy scale values. When these two are integer ratios, the~phase-space web is distinct, with~clear allowed and forbidden regions. By~contrast, when the two are irrational, the~web structure collapses, allowing the entire phase space. Floquet systems show that the discrete energy states $\groundom$ endure through time evolution, whereas the continuous states defined by $\fundom$ disperse. 

This scenario underscores the distinction between oscillator frequency $\groundom$ and~perturbation frequency $\fundom$ \cite{ENGEL2007}. We focused on the "cyclotron resonance'' case in Figure~\ref{fig:Hw_impulses_2} and the subsequent graphs, where $\groundom=\fundom$. Here, the~periodic kicks of the Floquet system coincided with the natural motion of the unperturbed~oscillator.

%In \cite{ENGEL2007}, this situation (and the case where these frequencies are integer multiples of each other) is explored and shown to relate to structured stochastic webs in phase space. 

%When the perturbing potential is zero, the amplitude for transitioning to each allowable oscillator state is zero, since the $sinc$ is zero precisely at multiples of the oscillator frequency. Increasing the potential moves the zeros off the axis, so they are still minima, but non-zero. (see figs. $\ref{fig:methodcompare-spectra-1}$- $\ref{fig:methodcompare-spectra-3}$). Due to the perturbation, the allowable transitions of the harmonic oscillator become more likely.
%Notice the intermediate states (l or b) go over all states, not just positive.
%Eqn 2.38 http://www.m-engel.de/m_engel_diss_2ol_l2h/node21.html
%mm'->az
%l->b

\subsection{Frequency Sampling~Interpretation} \label{hdr:sampling}

It is interesting to note the similarity between Equation \eqref{eqn:mainResult1} and the Shannon--Nyquist sampling theorem
\begin{align}
    \label{eqn:shannon}
    \begin{split}    
    f(t) = \sum_i f(n \DeltaT) \cdot sinc(\om t - n \pi).
    \end{split}
\end{align}

The theorem states that any signal $f(t)$ can be exactly reconstructed from its samples, $f(n \DeltaT)$, using a series of ideal $sinc$ interpolation functions centered on each sample, separated by a Nyquist~period.

In the case of Equation \eqref{eqn:mainResult1}, a~signal in the frequency domain, $c(\om)$, is smoothly reconstructed from discrete samples, $c(n \Delta \groundom)$, using a series of $sinc$-like interpolation functions centered on each energy eigenstate. In~the case of zero potential ($\tilde{V}(\om)=\delta(\om)$), if~we constrain the measurement window to be the inverse of the oscillator frequency, $\Time = 1/\fundom = 1/\groundom$, then Equation \eqref{eqn:mainResult1} exactly reconstructs the original wavefunction with an ideal $sinc$ interpolator. As~the perturbation increases from zero, the~interpolation function changes, and~the reconstructed signal is no longer identical to the~original.

%All appendix sections must be cited in the main text. In the appendices, Figures, Tables, etc. should be labeled, starting with ``A''---e.g., Figure A1, Figure A2, etc.

%%%%%%%%%%%%%%%%%%%%%%%%%%%%%%%%%%%%%%%%%%
%\begin{adjustwidth}{-\extralength}{0cm}

\begin{adjustwidth}{-\extralength}{0cm}
%\centering %% If there is a figure in wide page, please release command \centering
\reftitle{References}

% Please provide either the correct journal abbreviation (e.g., according to the “List of Title Word Abbreviations” http://www.issn.org/services/online-services/access-to-the-ltwa/) or the full name of the journal.

%\bibliography{main.bib}

\begin{thebibliography}{999}

\bibitem[Schrödinger(1926e)]{Schrod1926e}
Schrödinger, E.
\newblock Quantisierung als Eigenwertproblem (Vierte Mitteilung).
\newblock {\em Ann. Phys.} {\bf 1926e}, {\em 81}.

\bibitem[Dirac(1930)]{DIRAC1930}
Dirac, P.A.M.
\newblock {\em The Principles of Quantum Mechanics}; Oxford: Clarendon Press,  1930.

\bibitem[Dyson(1952)]{DYSON1952}
Dyson, F.J.
\newblock Divergence of Perturbation Theory in Quantum Electrodynamics.
\newblock {\em Phys. Rev.} {\bf 1952}, {\em 85},~631--632.
\newblock {\url{https://doi.org/10.1103/PhysRev.85.631}}.

\bibitem[Schwartz(2014)]{SCHWARTZ2014}
Schwartz, M.D.
\newblock {\em Quantum Field Theory and the Standard Model}; Cambridge University Press,  2014.

\bibitem[Walker and Gathright(1994)]{WALKER1994}
Walker, J.S.; Gathright, J.
\newblock Exploring one-dimensional quantum mechanics with transfer matrices.
\newblock {\em Am. J. Phys.} {\bf 1994}, {\em 62}.
\newblock {\url{https://doi.org/10.1119/1.17541}}.

\bibitem[Feynman(1948)]{FEYNMAN1948}
Feynman, R.P.
\newblock Space-Time Approach to Non-Relativistic Quantum Mechanics.
\newblock {\em Rev. Mod. Phys.} {\bf 1948}, {\em 20},~367--387.

\bibitem[Feynman(1948b{\natexlab{a}})]{FEYNMAN1948b}
Feynman, R.P.
\newblock A Relativistic Cut-off for Classical Electrodynamics.
\newblock {\em Phys. Rev.} {\bf 1948b}, {\em 74},~939--46.

\bibitem[Feynman(1948b{\natexlab{b}})]{FEYNMAN1948c}
Feynman, R.P.
\newblock Relativistic Cut-off for Quantum Electrodynamics.
\newblock {\em Phys. Rev.} {\bf 1948b}, {\em 74},~1430--38.

\bibitem[Schroeter(2018)]{GRIFFITHS2018}
Schroeter, D.J.G.D.F.
\newblock {\em Introduction to Quantum Mechanics, 3rd Edition}; University Cambridge Press,  2018.

\bibitem[Paganin and Pelliccia(2021)]{PAGANIN2021}
Paganin, D.M.; Pelliccia, D.
\newblock X-ray phase-contrast imaging: a broad overview of some fundamentals.
\newblock {\em Adv. Imaging Electron Phys.} {\bf 2021}, {\em 218},~63--158.

\bibitem[Strang(1968)]{STRANG1968}
Strang, G.
\newblock On the construction and comparison of difference schemes.
\newblock {\em SIAM Journal on Numerical Analysis} {\bf 1968}, {\em 5},~506--517.

\bibitem[Kosloff and Kosloff(1983)]{KOSLOFF1983}
Kosloff, D.; Kosloff, R.
\newblock A Fourier Method Solution for the Time Dependent Schr\"{o}dinger Equation as a Tool in Molecular Dynamics.
\newblock {\em Journal of Computational Physics} {\bf 1983}, {\em 52},~35--53.

\bibitem[Dateo et~al.(1991)Dateo, Engel, Almeida, and Horia]{DATEO1991}
Dateo, C.E.; Engel, V.; Almeida, R.; Horia, M.
\newblock Numerical solutions of the time-dependent Schrödinger equation in spherical coordinates by Fourier transform methods.
\newblock {\em Computer Physics Communications} {\bf 1991}, {\em 63},~435--445.

\bibitem[Van~Dyck(1985)]{VANDYCK1985}
Van~Dyck, D.
\newblock Image Calculations in High-Resolution Electron Microscopy: Problems. Progress. and Prospects.
\newblock {\em Adv. Electron. Electron. Phys.} {\bf 1985}, {\em 65},~295--355.

\bibitem[{Taha} and {Ablowitz}(1984)]{TAHA1984}
{Taha}, T.R.; {Ablowitz}, M.I.
\newblock {Analytical and Numerical Aspects of Certain Nonlinear Evolution Equations. II. Numerical, Nonlinear Schr{\"o}dinger Equation}.
\newblock {\em Journal of Computational Physics} {\bf 1984}, {\em 55},~203--230.
\newblock {\url{https://doi.org/10.1016/0021-9991(84)90003-2}}.

\bibitem[Bandrauk and Shen(1992)]{BANDRAUK1992}
Bandrauk, A.D.; Shen, H.
\newblock Higher order exponential split operator method for solving time-dependent Schrodinger equations.
\newblock {\em Can. J. Chem} {\bf 1992}, {\em 70}.

\bibitem[Hansson and Wabnitz(2016)]{HANSSON2016}
Hansson, T.; Wabnitz, S.
\newblock Dynamics of microresonator frequency comb generation: Models and stability.
\newblock {\em Nanophotonics} {\bf 2016}, {\em 5}.
\newblock {\url{https://doi.org/10.1515/nanoph-2016-0012}}.

\bibitem[Nelson-Isaacs(2021)]{NELSONISAACS2021}
Nelson-Isaacs, S.E.
\newblock Spacetime Paths as a Whole.
\newblock {\em Quantum reports} {\bf 2021}, {\em 3},~13--41.
\newblock {\url{https://doi.org/10.3390/quantum3010002}}.

\bibitem[Hong et~al.(1987)Hong, Ou, and Mandel]{HONG1987}
Hong, C.K.; Ou, Z.Y.; Mandel, L.
\newblock Measurement of subpicosecond time intervals between two photons by interference.
\newblock {\em Phys. Rev. Lett.} {\bf 1987}, {\em 59},~2044--2046.
\newblock {\url{https://doi.org/10.1103/PhysRevLett.59.2044}}.

\bibitem[Davis et~al.(2018)Davis, Thiel, Karpi\ifmmode~\acute{n}\else \'{n}\fi{}ski, and Smith]{DAVIS2018}
Davis, A.O.C.; Thiel, V.; Karpi\ifmmode~\acute{n}\else \'{n}\fi{}ski, M.; Smith, B.J.
\newblock Measuring the Single-Photon Temporal-Spectral Wave Function.
\newblock {\em Phys. Rev. Lett.} {\bf 2018}, {\em 121},~083602.
\newblock {\url{https://doi.org/10.1103/PhysRevLett.121.083602}}.

\bibitem[Mosley et~al.(2008)Mosley, Lundeen, Smith, Wasylczyk, U'Ren, Silberhorn, and Walmsley]{MOSLEY2008}
Mosley, P.J.; Lundeen, J.S.; Smith, B.J.; Wasylczyk, P.; U'Ren, A.B.; Silberhorn, C.; Walmsley, I.A.
\newblock Heralded Generation of Ultrafast Single Photons in Pure Quantum States.
\newblock {\em Phys. Rev. Lett.} {\bf 2008}, {\em 100},~133601.
\newblock {\url{https://doi.org/10.1103/PhysRevLett.100.133601}}.

\bibitem[M\"uller et~al.(2017)M\"uller, Tentrup, Bienert, Morigi, and Eschner]{MULLER2017}
M\"uller, P.; Tentrup, T.; Bienert, M.; Morigi, G.; Eschner, J.
\newblock Spectral properties of single photons from quantum emitters.
\newblock {\em Phys. Rev. A} {\bf 2017}, {\em 96},~023861.
\newblock {\url{https://doi.org/10.1103/PhysRevA.96.023861}}.

\bibitem[Tamma and Laibacher(2015)]{TAMMALAIBACHER2015}
Tamma, V.; Laibacher, S.
\newblock Multiboson Correlation Interferometry with Arbitrary Single-Photon Pure States.
\newblock {\em Physical Review Letters} {\bf 2015}, {\em 114},~243601.
\newblock {\url{https://doi.org/10.1103/PhysRevLett.114.243601}}.

\bibitem[Laibacher and Tamma(2018)]{TAMMALAIBACHER2018}
Laibacher, S.; Tamma, V.
\newblock Symmetries and entanglement features of inner-mode-resolved correlations of interfering nonidentical photons.
\newblock {\em Phys. Rev. A} {\bf 2018}, {\em 98},~053829.
\newblock {\url{https://doi.org/10.1103/PhysRevA.98.053829}}.

\bibitem[Tamma and Laibacher(2021)]{TAMMA2021}
Tamma, V.; Laibacher, S.
\newblock Boson sampling with random numbers of photons.
\newblock {\em Phys. Rev. A} {\bf 2021}, {\em 104},~032204.
\newblock {\url{https://doi.org/10.1103/PhysRevA.104.032204}}.

\bibitem[Tamma and Laibacher(2023)]{Tamma2023}
Tamma, V.; Laibacher, S.
\newblock Scattershot multiboson correlation sampling with random photonic inner-mode multiplexing.
\newblock {\em The European Physical Journal Plus} {\bf 2023}, {\em 138},~335.
\newblock {\url{https://doi.org/10.1140/epjp/s13360-023-03941-2}}.

\bibitem[Wang et~al.(2018)Wang, Jing, Sun, Yang, Yu, Tamma, Bao, and Pan]{wang2018experimental}
Wang, X.J.; Jing, B.; Sun, P.F.; Yang, C.W.; Yu, Y.; Tamma, V.; Bao, X.H.; Pan, J.W.
\newblock Experimental Time-Resolved Interference with Multiple Photons of Different Colors.
\newblock {\em Phys. Rev. Lett.} {\bf 2018}, {\em 121},~080501.
\newblock {\url{https://doi.org/10.1103/PhysRevLett.121.080501}}.

\bibitem[Triggiani et~al.(2023)Triggiani, Psaroudis, and Tamma]{TRIGGIANI2023}
Triggiani, D.; Psaroudis, G.; Tamma, V.
\newblock Ultimate Quantum Sensitivity in the Estimation of the Delay between two Interfering Photons through Frequency-Resolving Sampling.
\newblock {\em Physical Review Applied} {\bf 2023}, {\em 19},~044068.
\newblock {\url{https://doi.org/10.1103/PhysRevApplied.19.044068}}.

\bibitem[Cui et~al.(2012)Cui, Li, and Zhao]{Cui2012}
Cui, L.; Li, X.; Zhao, N.
\newblock Minimizing the frequency correlation of photon pairs in photonic crystal fibers.
\newblock {\em New Journal of Physics} {\bf 2012}, {\em 14},~123001.
\newblock {\url{https://doi.org/10.1088/1367-2630/14/12/123001}}.

\bibitem[Zhang et~al.(2024)Zhang, Sun, Hirschman, Shariatdoust, Belli, and Carbajo]{Zhang2024Optimizing}
Zhang, H.; Sun, L.; Hirschman, J.; Shariatdoust, M.S.; Belli, F.; Carbajo, S.
\newblock Optimizing Spectral Phase Transfer in Four-Wave Mixing with Gas-filled Capillaries: A Trade-off Study.
\newblock {\em arXiv:2404.16993 [physics.optics]} {\bf 2024}.

\bibitem[Asavanant and Furusawa(2024)]{Asavanant2024Multipartite}
Asavanant, W.; Furusawa, A.
\newblock Multipartite continuous-variable optical quantum entanglement: Generation and application.
\newblock {\em Physical Review A} {\bf 2024}, {\em 109},~040101.
\newblock {\url{https://doi.org/10.1103/PhysRevA.109.040101}}.

\bibitem[Bartolucci et~al.(2023)Bartolucci, Birchall, Bomb{\'i}n, Cable, Dawson, Gimeno-Segovia, Johnston, Kieling, Nickerson, Pant, Pastawski, Rudolph, and Sparrow]{Bartolucci2023}
Bartolucci, S.; Birchall, P.; Bomb{\'i}n, H.; Cable, H.; Dawson, C.; Gimeno-Segovia, M.; Johnston, E.; Kieling, K.; Nickerson, N.; Pant, M.;  et~al.
\newblock Fusion-based quantum computation.
\newblock {\em Nature Communications} {\bf 2023}, {\em 14},~912.
\newblock {\url{https://doi.org/10.1038/s41467-023-36493-1}}.

\bibitem[Lu et~al.(2024)Lu, Krasavin, Lan, Zayats, and Dai]{Lu2024}
Lu, W.; Krasavin, A.V.; Lan, S.; Zayats, A.V.; Dai, Q.
\newblock Gradient-induced long-range optical pulling force based on photonic band gap.
\newblock {\em Light: Science \& Applications} {\bf 2024}, {\em 13},~93.
\newblock {\url{https://doi.org/10.1038/s41377-024-01452-y}}.

\bibitem[Neuman and Block(2004)]{NEUMAN2004}
Neuman, K.C.; Block, S.M.
\newblock Optical trapping.
\newblock {\em Review of Scientific Instruments} {\bf 2004}, {\em 75},~2787--2809,  \href{http://xxx.lanl.gov/abs/https://pubs.aip.org/aip/rsi/article-pdf/75/9/2787/19215937/2787\_1\_online.pdf}{{\normalfont [https://pubs.aip.org/aip/rsi/article-pdf/75/9/2787/19215937/2787\_1\_online.pdf]}}.
\newblock {\url{https://doi.org/10.1063/1.1785844}}.

\bibitem[Pérez-García et~al.(2023)Pérez-García, Selin, Ciarlo, Magazzù, Pesce, Sasso, Volpe, Pérez~Castillo, and Arzola]{Perez-Garcia2023}
Pérez-García, L.; Selin, M.; Ciarlo, A.; Magazzù, A.; Pesce, G.; Sasso, A.; Volpe, G.; Pérez~Castillo, I.; Arzola, A.V.
\newblock Optimal calibration of optical tweezers with arbitrary integration time and sampling frequencies: a general framework [Invited].
\newblock {\em Biomed Opt Express} {\bf 2023}, {\em 14},~6442--6469.
\newblock [Invited], {\url{https://doi.org/10.1364/BOE.495468}}.

\bibitem[Panda and Benjamin(2024)]{Panda2024Quantum}
Panda, D.K.; Benjamin, C.
\newblock Quantum cryptographic protocols with dual messaging system via 2D alternate quantum walks and genuine single particle entangled states.
\newblock {\em arXiv:2405.00663 [quant-ph]} {\bf 2024}.

\bibitem[Lounis(2014)]{LOUNIS2014}
Lounis, S.
\newblock Theory of Scanning Tunneling Microscopy {\bf 2014}.

\bibitem[Gottlieb and Wesoloski(2006)]{GOTTLIEB2006}
Gottlieb, A.D.; Wesoloski, L.
\newblock Bardeen’s Tunneling Theory as Applied to Scanning Tunneling Microscopy: A Technical Guide to the Traditional Interpretation.
\newblock {\em Nanotechnology} {\bf 2006}, {\em 17}.
\newblock {\url{https://doi.org/10.1088/0957-4484/17/8/R01}}.

\bibitem[Grewal et~al.(2024)Grewal, Leon, Kuhnke, Kern, and Gunnarsson]{Grewal2024}
Grewal, A.; Leon, C.C.; Kuhnke, K.; Kern, K.; Gunnarsson, O.
\newblock Scanning Tunneling Microscopy for Molecules: Effects of Electron Propagation into Vacuum.
\newblock {\em ACS Nano} {\bf 2024}.
\newblock {\url{https://doi.org/10.1021/acsnano.3c12315}}.

\bibitem[Dessai and Kulkarni(2022)]{DessaiKulkarni2022}
Dessai, M.; Kulkarni, A.V.
\newblock Calculation of tunneling current across trapezoidal potential barrier in a scanning tunneling microscope.
\newblock {\em Journal of Applied Physics} {\bf 2022}, {\em 132},~244901.
\newblock {\url{https://doi.org/10.1063/5.0132208}}.

\bibitem[Gaida et~al.(2024)Gaida, Lourenço-Martins, Sivis, Rittmann, Feist, de~Abajo, and Ropers]{Gaida2024}
Gaida, J.H.; Lourenço-Martins, H.; Sivis, M.; Rittmann, T.; Feist, A.; de~Abajo, F.J.G.; Ropers, C.
\newblock Attosecond electron microscopy by free-electron homodyne detection.
\newblock {\em Nature Photonics} {\bf 2024}, {\em 18},~509--515.
\newblock {\url{https://doi.org/10.1038/s41566-024-01380-8}}.

\bibitem[Cao et~al.(2024)Cao, Eckner, Yelin, Young, Jandura, Yan, Kim, Pupillo, Ye, Oppong, and Kaufman]{cao2024multiqubit}
Cao, A.; Eckner, W.J.; Yelin, T.L.; Young, A.W.; Jandura, S.; Yan, L.; Kim, K.; Pupillo, G.; Ye, J.; Oppong, N.D.;  et~al.
\newblock Multi-qubit gates and 'Schr\"odinger cat' states in an optical clock,  2024,  \href{http://xxx.lanl.gov/abs/2402.16289}{{\normalfont [arXiv:quant-ph/2402.16289]}}.

\bibitem[Kawasaki(2024{\natexlab{a}})]{Kawasaki2024}
Kawasaki, A.
\newblock Real-time observation of picosecond-timescale optical quantum entanglement toward ultrafast quantum information processing.
\newblock {\em arXiv} {\bf 2024},  \href{http://xxx.lanl.gov/abs/2403.07357}{{\normalfont [arXiv:quant-ph/2403.07357]}}.

\bibitem[Kawasaki(2024{\natexlab{b}})]{Kawasaki2024HighRate}
Kawasaki, A.
\newblock High-rate Generation and State Tomography of Non-Gaussian Quantum States for Ultra-fast Clock Frequency Quantum Processors.
\newblock {\em arXiv} {\bf 2024},  \href{http://xxx.lanl.gov/abs/2402.17408}{{\normalfont [arXiv:quant-ph/2402.17408]}}.

\bibitem[Nishidome et~al.(2024)Nishidome, Omoto, Nagai, Uchida, Murakami, Eda, Okubo, Ueji, Yomogida, Kono, Tanaka, and Yanagi]{Nishidome2024}
Nishidome, H.; Omoto, M.; Nagai, K.; Uchida, K.; Murakami, Y.; Eda, J.; Okubo, H.; Ueji, K.; Yomogida, Y.; Kono, J.;  et~al.
\newblock Influence of Laser Intensity and Location of the Fermi Level on Tunneling Processes for High-Harmonic Generation in Arrayed Semiconducting Carbon Nanotubes.
\newblock {\em ACS Photonics} {\bf 2024}, {\em 11},~171--179.
\newblock {\url{https://doi.org/10.1021/acsphotonics.3c01244}}.

\bibitem[Majidi et~al.(2023)Majidi, Aghbolaghi, Navid, and Mokhlesi]{Majidi2023}
Majidi, S.; Aghbolaghi, R.; Navid, H.A.; Mokhlesi, R.
\newblock Optimization of cut-off frequency in high harmonic generation in noble gas.
\newblock {\em Applied Physics B} {\bf 2023}, {\em 130},~11.
\newblock {\url{https://doi.org/10.1007/s00340-023-08139-z}}.

\bibitem[Farkas and T{\'o}th(1992)]{Farkas1992}
Farkas, G.; T{\'o}th, C.
\newblock Proposal for attosecond light pulse generation using laser-induced multiple harmonic conversion processes in rare gases.
\newblock {\em Physics Letters A} {\bf 1992}, {\em 168},~447--450.

\bibitem[Lewenstein et~al.(1994)Lewenstein, Balcou, Ivanov, L'Huillier, and Corkum]{LEWENSTEIN1994}
Lewenstein, M.; Balcou, P.; Ivanov, M.Y.; L'Huillier, A.; Corkum, P.B.
\newblock Theory of high-harmonic generation by low-frequency laser fields.
\newblock {\em Phys. Rev. A} {\bf 1994}, {\em 49},~2117--2132.
\newblock {\url{https://doi.org/10.1103/PhysRevA.49.2117}}.

\bibitem[Ryabikin et~al.(2023)Ryabikin, Emelin, and Strelkov]{Ryabikin:2023}
Ryabikin, M.Y.; Emelin, M.Y.; Strelkov, V.V.
\newblock Attosecond electromagnetic pulses: generation, measurement, and application. Attosecond metrology and spectroscopy.
\newblock {\em Phys. Usp.} {\bf 2023}, {\em 66},~360--380.
\newblock {\url{https://doi.org/10.3367/UFNe.2021.10.039078}}.

\bibitem[Hänsch(1990)]{HANSCH1990}
Hänsch, T.
\newblock A proposed sub-femtosecond pulse synthesizer using separate phase-locked laser oscillators.
\newblock {\em Optics Communications} {\bf 1990}, {\em 80},~71--75.
\newblock {\url{https://doi.org/https://doi.org/10.1016/0030-4018(90)90509-R}}.

\bibitem[Abele et~al.(2010)Abele, Jenke, Leeb, and Schmiedmayer]{ABELE2010}
Abele, H.; Jenke, T.; Leeb, H.; Schmiedmayer, J.
\newblock Ramsey's method of separated oscillating fields and its application to gravitationally induced quantum phase shifts.
\newblock {\em Phys. Rev. D} {\bf 2010}, {\em 81},~065019.
\newblock {\url{https://doi.org/10.1103/PhysRevD.81.065019}}.

\bibitem[Jenke et~al.(2011)Jenke, Geltenbort, Lemmel, and Abele]{Jenke2011}
Jenke, T.; Geltenbort, P.; Lemmel, H.; Abele, H.
\newblock Realization of a gravity-resonance-spectroscopy technique.
\newblock {\em Nature Physics} {\bf 2011}, {\em 7},~468--472.
\newblock {\url{https://doi.org/10.1038/nphys1970}}.

\bibitem[Villegas-Martínez et~al.(2016)Villegas-Martínez, Soto-Eguibar, and Moya-Cessa]{VILLEGAS2016}
Villegas-Martínez, B.M.; Soto-Eguibar, F.; Moya-Cessa, H.M.
\newblock Application of Perturbation Theory to a Master Equation.
\newblock {\em Advances in Mathematical Physics} {\bf 2016}, {\em 2016},~9265039.
\newblock {\url{https://doi.org/https://doi.org/10.1155/2016/9265039}}.

\bibitem[Krijtenburg-Lewerissa et~al.(2017)Krijtenburg-Lewerissa, Pol, Brinkman, and van Joolingen]{Krijtenburg2017}
Krijtenburg-Lewerissa, K.; Pol, H.J.; Brinkman, A.; van Joolingen, W.R.
\newblock Insights into teaching quantum mechanics in secondary and lower undergraduate education.
\newblock {\em Phys. Rev. Phys. Educ. Res.} {\bf 2017}, {\em 13},~010109.
\newblock {\url{https://doi.org/10.1103/PhysRevPhysEducRes.13.010109}}.

\bibitem[Singh et~al.(2006)Singh, Belloni, and Christian]{Singh2006}
Singh, C.; Belloni, M.; Christian, W.
\newblock {Improving students’ understanding of quantum mechanics}.
\newblock {\em Physics Today} {\bf 2006}, {\em 59},~43--49,  \href{http://xxx.lanl.gov/abs/https://pubs.aip.org/physicstoday/article-pdf/59/8/43/10971886/43\_1\_online.pdf}{{\normalfont [https://pubs.aip.org/physicstoday/article-pdf/59/8/43/10971886/43\_1\_online.pdf]}}.
\newblock {\url{https://doi.org/10.1063/1.2349732}}.

\bibitem[Nodurft et~al.(2022)Nodurft, Shaw, Glasser, Kirby, and Searles]{Nodurft2022}
Nodurft, I.C.; Shaw, H.C.; Glasser, R.T.; Kirby, B.T.; Searles, T.A.
\newblock Generation of polarization entanglement via the quantum Zeno effect.
\newblock {\em Opt. Express} {\bf 2022}, {\em 30},~31971--31985.
\newblock {\url{https://doi.org/10.1364/OE.464550}}.

\bibitem[Sakurai and Napolitano(2011)]{SakuraiNapolitano2011}
Sakurai, J.J.; Napolitano, J.
\newblock {\em Modern Quantum Mechanics}, 2nd ed.; Pearson Education Inc,  2011.

\bibitem[Tokmakoff(2014)]{Tokmakoff2014}
Tokmakoff, A.
\newblock Time-Dependent Quantum Mechanics and Spectroscopy.
\newblock University of Chicago,  2014.
\newblock Available at: \url{https://tdqms.uchicago.edu/full-tdqms-notes-upload-december-2014/}.

\bibitem[J.~M.~Zhang(2016)]{ZHANG2016}
J.~M.~Zhang, Y.L.
\newblock Fermi's golden rule: its derivation and breakdown by an ideal model.
\newblock {\em Eur. J. Phys.} {\bf 2016}, {\em 37}.
\newblock {\url{https://doi.org/https://doi.org/10.1088/0143-0807/37/6/065406}}.

\bibitem[Baym(1969)]{baym1969}
Baym, G.
\newblock {\em Lectures On Quantum Mechanics}, 1 ed.; CRC Press,  1969.
\newblock {\url{https://doi.org/10.1201/9780429499265}}.

\bibitem[Fowler(2007)]{Fowler2007}
Fowler, M.
\newblock Time-Dependent Perturbation Theory,  2007.
\newblock Accessed: 2023-12-26.

\bibitem[Tamma and Laibacher(2016)]{TAMMA2016BOSON}
Tamma, V.; Laibacher, S.
\newblock Boson sampling with non-identical single photons.
\newblock {\em Journal of Modern Optics} {\bf 2016}, {\em 63},~41--45.

\bibitem[Li et~al.(2008)Li, Ma, Ou, Yang, Cui, and Yu]{Li2008}
Li, X.; Ma, X.; Ou, Z.Y.; Yang, L.; Cui, L.; Yu, D.
\newblock Spectral study of photon pairs generated in dispersion shifted fiber with a pulsed pump.
\newblock {\em Opt. Express} {\bf 2008}, {\em 16},~32--44.
\newblock {\url{https://doi.org/10.1364/OE.16.000032}}.

\bibitem[Chen et~al.(2005)Chen, Li, and Kumar]{CHEN2005}
Chen, J.; Li, X.; Kumar, P.
\newblock Two-photon-state generation via four-wave mixing in optical fibers.
\newblock {\em Phys. Rev. A} {\bf 2005}, {\em 72},~033801.
\newblock {\url{https://doi.org/10.1103/PhysRevA.72.033801}}.

\bibitem[Sharping et~al.(2004)Sharping, Chen, Li, Kumar, and Windeler]{SHARPING2004}
Sharping, J.E.; Chen, J.; Li, X.; Kumar, P.; Windeler, R.S.
\newblock Quantum-correlated twin photons from microstructure fiber.
\newblock {\em Opt. Express} {\bf 2004}, {\em 12},~3086--3094.
\newblock {\url{https://doi.org/10.1364/OPEX.12.003086}}.

\bibitem[Garay-Palmett et~al.(2007)Garay-Palmett, McGuinness, Cohen, Lundeen, Rangel-Rojo, U'Ren, Raymer, McKinstrie, Radic, and Walmsley]{GARAYPALMETT2007}
Garay-Palmett, K.; McGuinness, H.J.; Cohen, O.; Lundeen, J.S.; Rangel-Rojo, R.; U'Ren, A.B.; Raymer, M.G.; McKinstrie, C.J.; Radic, S.; Walmsley, I.A.
\newblock Photon pair-state preparation with tailored spectral properties by spontaneous four-wave mixing in photonic-crystal fiber.
\newblock {\em Opt. Express} {\bf 2007}, {\em 15},~14870--14886.
\newblock {\url{https://doi.org/10.1364/OE.15.014870}}.

\bibitem[Keller and Rubin(1997)]{KELLER1997}
Keller, T.E.; Rubin, M.H.
\newblock Theory of two-photon entanglement for spontaneous parametric down-conversion driven by a narrow pump pulse.
\newblock {\em Phys. Rev. A} {\bf 1997}, {\em 56},~1534--1541.
\newblock {\url{https://doi.org/10.1103/PhysRevA.56.1534}}.

\bibitem[Rubin et~al.(1994)Rubin, Klyshko, Shih, and Sergienko]{RUBIN1994}
Rubin, M.H.; Klyshko, D.N.; Shih, Y.H.; Sergienko, A.V.
\newblock Theory of two-photon entanglement in type-II optical parametric down-conversion.
\newblock {\em Phys. Rev. A} {\bf 1994}, {\em 50},~5122--5133.
\newblock {\url{https://doi.org/10.1103/PhysRevA.50.5122}}.

\bibitem[Erez et~al.(2004)Erez, Aharonov, Reznik, and Vaidman]{EREZ2004}
Erez, N.; Aharonov, Y.; Reznik, B.; Vaidman, L.
\newblock Correcting quantum errors with the Zeno effect.
\newblock {\em Phys. Rev. A} {\bf 2004}, {\em 69},~062315.
\newblock {\url{https://doi.org/10.1103/PhysRevA.69.062315}}.

\bibitem[Franson et~al.(2004)Franson, Jacobs, and Pittman]{FRANSON2004}
Franson, J.D.; Jacobs, B.C.; Pittman, T.B.
\newblock Quantum computing using single photons and the Zeno effect.
\newblock {\em Phys. Rev. A} {\bf 2004}, {\em 70},~062302.
\newblock {\url{https://doi.org/10.1103/PhysRevA.70.062302}}.

\bibitem[Bayındır et~al.(2021)Bayındır, Altintas, and Ozaydin]{BAYINDIR2021}
Bayındır, C.; Altintas, A.A.; Ozaydin, F.
\newblock Self-localized solitons of a q-deformed quantum system.
\newblock {\em Communications in Nonlinear Science and Numerical Simulation} {\bf 2021}, {\em 92},~105474.
\newblock {\url{https://doi.org/https://doi.org/10.1016/j.cnsns.2020.105474}}.

\bibitem[Engel(2007)]{ENGEL2007}
Engel, U.M.
\newblock {\em On Quantum Chaos, Stochastic Webs and Localization in a Quantum Mechanical Kick System}; Logos Verlag,  2007.

\end{thebibliography}

\begin{thebibliography}{999}
% Reference 1
\bibitem[Author1(year)]{ref-journal}
Author~1, T. The title of the cited article. {\em J. Abbr.} {\bf 2008}, {\em 10}, 142--149.
% Reference 2
\bibitem[Author2(year)]{ref-book1}
Author~2, L. The title of the cited contribution. In {\em The Book Title}; Editor 1, F., Editor 2, A., Eds.; Publishing House: City, Country, 2007; pp. 32--58.
% Reference 3
\bibitem[Author3(year)]{ref-book2}
Author 1, A.; Author 2, B. \textit{Book Title}, 3rd ed.; Publisher: Publisher Location, Country, 2008; pp. 154--196.
% Reference 4
\bibitem[Author4(year)]{ref-unpublish}
Author 1, A.B.; Author 2, C. Title of Unpublished Work. \textit{Abbreviated Journal Name} year, \textit{phrase indicating stage of publication (submitted; accepted; in press)}.
% Reference 5
\bibitem[Author5(year)]{ref-communication}
Author 1, A.B. (University, City, State, Country); Author 2, C. (Institute, City, State, Country). Personal communication, 2012.
% Reference 6
\bibitem[Author6(year)]{ref-proceeding}
Author 1, A.B.; Author 2, C.D.; Author 3, E.F. Title of presentation. In Proceedings of the Name of the Conference, Location of Conference, Country, Date of Conference (Day Month Year); Abstract Number (optional), Pagination (optional).
% Reference 7
\bibitem[Author7(year)]{ref-thesis}
Author 1, A.B. Title of Thesis. Level of Thesis, Degree-Granting University, Location of University, Date of Completion.
% Reference 8
\bibitem[Author8(year)]{ref-url}
Title of Site. Available online: URL (accessed on Day Month Year).
\end{thebibliography}

\PublishersNote{}
\end{adjustwidth}

%=====================================
% References, variant B: internal bibliography
%=====================================
\begin{comment}

\end{comment} 

% For the MDPI journals use author-date citation, please follow the formatting guidelines on http://www.mdpi.com/authors/references
% To cite two works by the same author: \citeauthor{ref-journal-1a} (\citeyear{ref-journal-1a}, \citeyear{ref-journal-1b}). This produces: Whittaker (1967, 1975)
% To cite two works by the same author with specific pages: \citeauthor{ref-journal-3a} (\citeyear{ref-journal-3a}, p. 328; \citeyear{ref-journal-3b}, p.475). This produces: Wong (1999, p. 328; 2000, p. 475)

%%%%%%%%%%%%%%%%%%%%%%%%%%%%%%%%%%%%%%%%%%

\end{document}